\pdfoutput=1
\documentclass[aps,prl,
twocolumn,
superscriptaddress,nobibnotes]{revtex4-1}

\usepackage{amsmath}
\usepackage{graphicx}
\usepackage{multirow}
\usepackage{color}
\usepackage{braket}

\usepackage{dcolumn}
\newcolumntype{.}{D{.}{.}{-1}}
\newcommand\eps{\varepsilon}

\usepackage{bm}

\renewcommand\vec[1]{\bm{\mathrm{#1}}}
\renewcommand{\thetable}{\arabic{table}}

\usepackage[explicit]{titlesec}

\usepackage[normalem]{ulem}

\usepackage[colorlinks=true,linkcolor=blue,citecolor=blue,urlcolor=blue]{hyperref}
\pdfpageattr{/Group << /S /Transparency /I true /CS /DeviceRGB>>}
\usepackage{chngcntr}
\newcounter{myseccnt}
\counterwithin*{figure}{myseccnt}
\counterwithin*{table}{myseccnt}

\begin{document}
\title{Performance of arsenene and antimonene double-gate MOSFETs from first principles}

\author{Giovanni Pizzi} 
  \altaffiliation{These authors contributed equally to this work}
  \affiliation{Theory and Simulation of
  Materials (THEOS) and National Centre for Computational Design and
  Discovery of Novel Materials (MARVEL), \'Ecole Polytechnique
  F\'ed\'erale de Lausanne, CH-1015 Lausanne, Switzerland}
\author{Marco Gibertini}
  \altaffiliation{These authors contributed equally to this work}
  \affiliation{Theory and Simulation of
  Materials (THEOS) and National Centre for Computational Design and
  Discovery of Novel Materials (MARVEL), \'Ecole Polytechnique
  F\'ed\'erale de Lausanne, CH-1015 Lausanne, Switzerland}
\author{Elias Dib} 
  \altaffiliation{These authors contributed equally to this work}
  \affiliation{Dipartimento di Ingegneria
  dell'Informazione, University of Pisa, 56122 Pisa, Italy}
\author{Nicola Marzari} \affiliation{Theory and Simulation of
  Materials (THEOS) and National Centre for Computational Design and
  Discovery of Novel Materials (MARVEL), \'Ecole Polytechnique
  F\'ed\'erale de Lausanne, CH-1015 Lausanne, Switzerland}
\author{Giuseppe Iannaccone} \affiliation{Dipartimento di Ingegneria
  dell'Informazione, University of Pisa, 56122 Pisa, Italy}
\author{Gianluca Fiori}
  \email{gfiori@mercurio.iet.unipi.it}
  \affiliation{Dipartimento di Ingegneria
  dell'Informazione, University of Pisa, 56122 Pisa, Italy}

\begin{abstract}
In the race towards high-performance ultra-scaled devices,
two-dimensional materials offer an alternative paradigm thanks to
their atomic thickness suppressing short-channel effects.
It is thus urgent to study the most promising candidates in realistic
configurations, and here we present detailed multiscale simulations
of field-effect transistors based on arsenene and antimonene monolayers as
channels. The accuracy of first-principles
approaches in describing electronic properties is combined
with the efficiency of tight-binding Hamiltonians based on maximally-localised
Wannier functions to compute the transport properties of the devices.
These simulations provide for the first time estimates on the upper limits
for the electron and hole mobilities in the Takagi's approximation,
including spin-orbit and multi-valley effects, and demonstrate that
ultra-scaled devices in the sub-10~nm scale show a performance that is
compliant with industry requirements. 
\end{abstract}

\maketitle
\setcounter{myseccnt}{1}

\section*{Introduction}
In the past decades the exponential increase in computing power
predicted by Moore's law has been enabled by scaling complementary metal-oxide-semiconductor (CMOS)
silicon-based devices, i.e., reducing their size and limiting at the
same time their power dissipation, while increasing the operating
frequency. With transistor dimensions going below 10~nm,
fundamental limitations are emerging both in terms of manufacturing
costs and device performance.  To sustain Moore's law, a paradigm shift either in device architecture or
in materials is needed.  

In this respect, using 2D systems as
conduction channels is definitely one of the most exciting
opportunities\cite{Dresselhaus2016}. Indeed, their ultimate thinness
can reduce short-channel effects, 
one of the main detrimental factors for devices at
ultrashort lengths.
For this reason, starting with the experimental realisation of 
graphene\cite{Novoselov2004},
single-layer materials have gained considerable attention for a large
number of different applications.  Several studies, in particular,
have targeted graphene as a component of novel devices, motivated by
its exciting electronic, mechanical and thermal properties, such as
its extremely high mobility~\cite{Novoselov2005}.  Despite its appeal,
graphene has regrettably no gap. Therefore, it is not suited for
electronic applications such as field-effect transistors (FETs), where
a semiconductor material with a finite gap is required for device
switching.
The first suitable candidate, the transition-metal dichalcogenide
(TMDC) MoS$_2$,\cite{Mak2010} has been shown to be an interesting
transistor material, even if its mobility is much lower than that of
graphene\cite{Radisavljevic2011}.  The list of relevant
two-dimensional systems has then been enriched by other
TMDCs~\cite{Tan2014} and by many other layered materials, such as
black phosphorus and its monolayer form phosphorene, that is promising for
its high mobility\cite{Li2014,Liu2014,Qiao2014,Han2014,Du2014,Avsar2015}.

In light of the current pace at which 2D materials are being
identified, we cannot expect that each new candidate is grown
experimentally with high quality and then devices with various geometries are 
fabricated, characterised and optimised.  
Instead, simulations can be used to efficiently determine and 
optimise materials properties and device
characteristics and filter only a few systems to send then to the laboratory.
Promising candidates can be considered, for instance, by
looking for materials chemically similar to existing ones. As an example,
two new monolayer materials composed of group--V elements (in analogy
with phosphorene) have been 
recently theoretically investigated:
arsenene and antimonene,\cite{Kamal2015,Wang2015,Shengli2015,Zhu2015} made of As and Sb, respectively. The authors
have put forward the hypothesis that they could be attractive for
device applications. While this suggestion is reasonable, only an
accurate simulation of a complete device can support this hypothesis
and will further stimulate experimental interest\cite{Lei2016} in these novel 2D materials.

This task is not straightforward, however, because the simulation of a
device requires a preliminary characterisation of the material. While
properties and parameters are available in the literature for
well-studied systems (such as bulk Si or III-V semiconductors), in
the case of new materials they are typically not available, nor they
can be easily extracted from known systems; they must instead be
calculated from first principles.  This
can be true even in simple cases: for instance, despite their chemical
similarity, arsenene and antimonene display very different electronic
and mechanical properties with respect to phosphorene, as they
originate from different allotropes and have thus a completely
different crystal structure.  On the other hand, performing a full
device simulation directly from first principles is computationally
out of reach.  Device simulators based on effective tight-binding
Hamiltonians~\cite{ViDES,ProcIEEE} are viable, but
require the knowledge of on-site and hopping energies, and a
few different methods have been proposed in the
literature to address the issue of bridging the different simulation 
scales~\cite{Porezag1995,Tan2013}. 

Here, we adopt
a multiscale approach based on maximally localised Wannier functions
(MLWF)~\cite{Marzari2012}, while providing a physical
understanding of the transport properties of monolayer As and Sb.
Basic electronic
properties are calculated from first principles using density-functional 
theory (DFT). The electronic wavefunctions are then used as
input to obtain MLWF in a multiscale approach, providing us with
an effective tight-binding Hamiltonian for the relevant electronic
bands around the fundamental gap, and retaining at the same time full
first-principles accuracy in the results.\cite{Lee2005,Bruzzone2014} 
MLWF are used to
characterise the material (e.g., effective masses) by exploiting
the efficient band interpolation, and as a localised tight-binding
basis set to simulate the currents in a complete device with a
non-equilibrium Green function (NEGF) formalism~\cite{Datta}. 
In particular, we
consider double-gate metal-oxide-semiconductor field-effect transistors (MOSFETs) based on arsenene and
antimonene channels and compare their performance against Industry
requirements.  We show that such devices have the potential to achieve
the target set by the International Technology Roadmap for
Semiconductors (ITRS)~\cite{ITRS} for future competitive devices for
high performance digital applications, in particular in terms of the
capability of behaving as
an outstanding switch even in the ultra-scaled regime.

\section*{Results}

\subsection*{Multiscale material characterisation}

The first step towards the multiscale simulation of devices based on
As and Sb monolayers is the computation from first principles of their
electronic structure. To perform this task, we carried out DFT
simulations using the Quantum ESPRESSO\cite{Giannozzi2009} suite of
codes, efficiently automated using AiiDA\cite{Pizzi2016} (more
details in the Methods section). All calculations reported here include
spin-orbit coupling (SOC) effects. In the Supplementary Note 1
we discuss in detail the effect of SOC and compare the results obtained here with those 
without SOC.

\begin{figure} [tbp]
\centering\includegraphics[width=8cm]{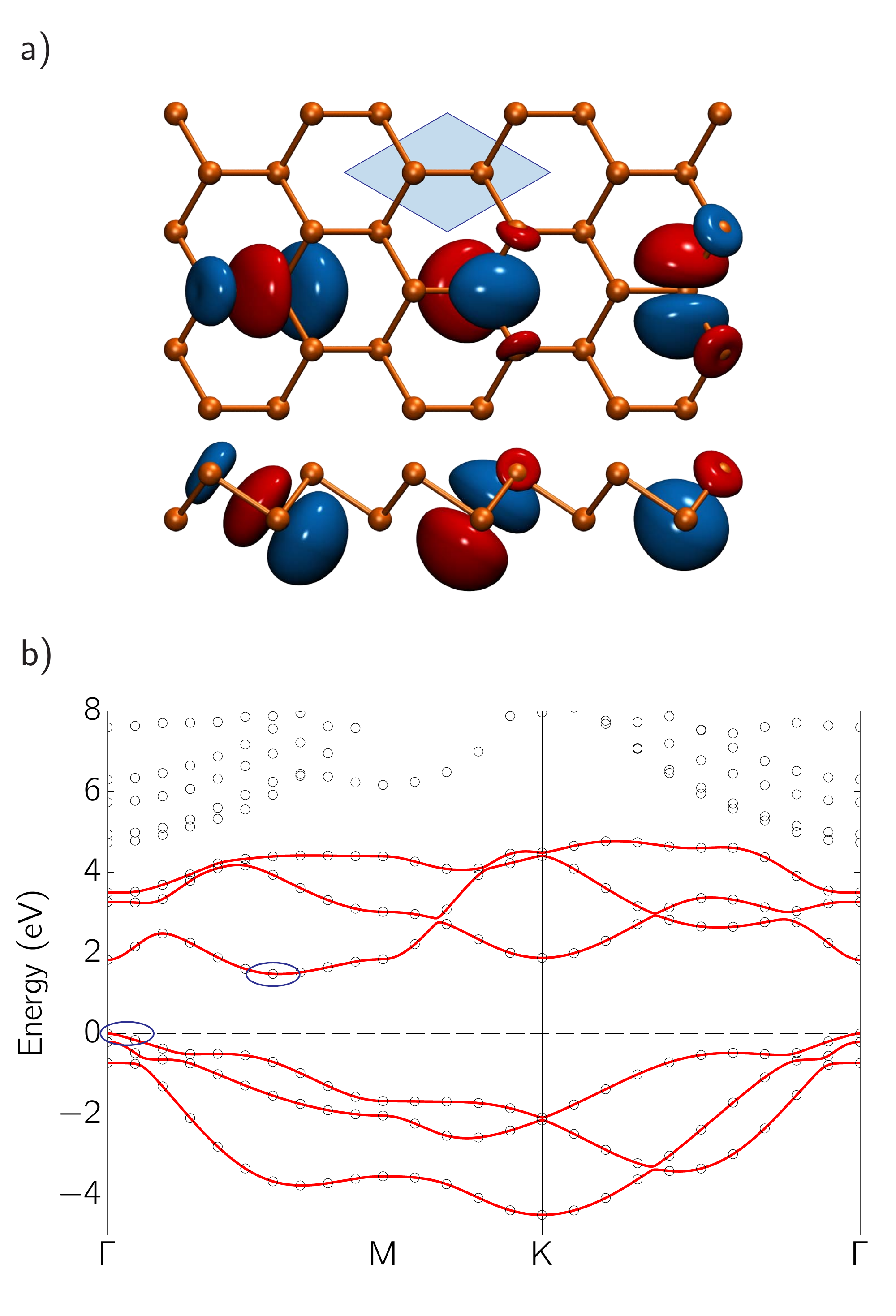}
\caption{\textbf{Structure, energy bands, and Wannier functions of As and Sb monolayers:} (a) Top and lateral views of As or Sb monolayers. The
  blue shaded region represents the primitive unit cell comprising two
  inequivalent atoms. The spatial profile of three maximally-localised
  Wannier functions is also reported. Isosurfaces of different colours
  (red and blue) correspond to opposite values of the real-valued
  Wannier functions.  (b) Energy bands of arsenene along a
  high-symmetry path in the Brillouin zone. Empty circles denote the
  results of a direct DFT calculation while red solid lines represent
  the Wannier-interpolated bands. Blue circles highlight the position
  of the valence band maximum and conduction band minimum. }
\label{fig:dft-wannier}
\end{figure}

In Fig.~\ref{fig:dft-wannier}a we
show the equilibrium crystal structure of arsenene and
antimonene. Differently to phosphorene, As and Sb monolayers are not
puckered, but display a buckled structure more similar to silicene or
germanene~\cite{Aufray2010,Davila2014}, with two inequivalent atoms
inside the primitive hexagonal unit cell lying on two different
planes. (Note that As and Sb have also been predicted to exist in a puckered structure,
but this phase is energetically less favourable~\cite{Kamal2015,Wang2015}).
The separation $d$ between the planes reads 1.394~\AA\ for As and
1.640~\AA\ for Sb, while the equilibrium lattice constant $a$ is,
respectively, 3.601~\AA\ and 4.122~\AA.  The band structure of both
materials is very similar and in Fig.~\ref{fig:dft-wannier}b we show
the energy bands along a high-symmetry path in the Brillouin zone
obtained from DFT (empty circles) in the case of arsenene (for the
bands of antimonene see Supplementary Figure 1). The DFT band gap
is indirect (for both materials)
and equal to $1.48$~eV for As and $1.00$~eV for Sb.
The maximum of the valence bands is located at the $\Gamma$ point and, without SOC,
it would be two-fold degenerate; the inclusion of the SOC splits the degeneracy 
(see Fig.~\ref{fig:dft-wannier}b and Supplementary Figure 2) and
the topmost valence band becomes non-degenerate, except for the twofold spin degeneracy.
The minimum of the conduction bands lies instead
along the $\Gamma-$M line and gives rise to six valleys.  

Further analyses of the electronic properties of arsenene and antimonene have
been performed by first mapping the Bloch eigenstates associated with
the bands around the gap into a set of maximally localised Wannier
functions\cite{Marzari2012}. We focused in particular on the three top
valence bands and three bottom conduction bands (per spin component). The main orbital
contribution to these bands comes from $p$-orbitals of the atoms that
form bonding and antibonding combinations around the gap. By
projecting over the $p$-orbitals of the two atoms in the primitive
cell, the standard localisation procedure leads to six Wannier
functions per spin component, three centred on one atom and three on the other. In
Fig.~\ref{fig:dft-wannier}a we show the spatial profile of the three
Wannier functions centred on atoms belonging to the lower plane (the
other three can be obtained simply by spatial inversion through a
mid-bond centre). They clearly have a $p$-like character with minor
contributions from neighbouring atoms. From the knowledge of these
Wannier functions it is straightforward to compute the matrix elements
of the Hamiltonian between them. 

In such a way, it becomes possible to
interpolate efficiently the Hamiltonian at any arbitrary $\vec{k}-$point in
reciprocal space, keeping the same accuracy of the underlying
first-principles simulation, but at an extremely reduced computational
cost.  In Fig.~\ref{fig:dft-wannier}b we show the
Wannier-interpolated energy bands (red solid lines) with a much denser
mesh than the original DFT results (empty circles) for As monolayer (bands for Sb monolayer
are shown in Supplementary Figure 1). Exploiting such
interpolation scheme, we also computed the effective masses for
relevant band extrema that crucially affect carrier mobilities and intraband tunnelling amplitudes. We both fitted the electronic bands along
principal directions and evaluated accurately the
density-of-states (DOS) on an extremely dense grid. The values of the
masses are reported in Table~\ref{tab:effmasses} for both
materials. In particular, for the valence band maxima, the SOC splits the degeneracy of
the two topmost bands at $\Gamma$. Since the magnitude of the splitting is 
large, for realistic band filling levels we can limit ourselves to consider only the topmost
valence band, with (isotropic) mass $m^{\rm v}$. (If SOC was neglected, we would need instead to
consider both degenerate bands, as discussed in Supplementary Notes 2--5 and Supplementary Figures 2--5.) For the conduction
bands, the isoenergies of the six valleys are oblate with a larger
effective mass $m^{\rm c}_{\rm L}$ in the longitudinal direction with respect to
the transverse effective mass $m^{\rm c}_{\rm T}$. The
effective DOS mass for each valley $m^{\rm c}_{\text{DOS}}$ computed
independently is in agreement with what can be expected from
geometrical arguments, i.e. $m^{\rm c}_{\text{DOS}} \approx
\sqrt{m^{\rm c}_{\text{L}}m^{\rm c}_{\text{T}}}$.

As a first assessment of the material properties toward the
realisation of a transistor device using arsenene or antimonene as
channel materials, we estimate 
whether the ballistic
approximation is valid at room-temperature ($T=300$\,K) 
in the ultrascaled sub-10~nm regime that
we investigate in this work. We will limit
the analysis only to the intrinsic
scattering with
longitudinal acoustic (LA) phonons.
As other scattering mechanisms may
be active in the system, the values that we calculate should be
considered as upper limits to the actual scattering times or,
equivalently, to the carrier mobility.
In particular, while out-of-plane (ZA) phonons may play
an important role in free-standing Dirac materials without planar symmetry~\cite{Fischetti2016,Gunst2016}, we do not consider
them here. While in a free-standing material scattering with
ZA phonons can be relevant, in our systems
the device geometry (presence of substrate and of top gates)
will shift the ZA phonon modes at finite energy, reducing their
impact on the mobilities.\cite{Slotman2014}

While accurate values for the electron-phonon scattering 
terms can be obtained fully ab-initio~\cite{Savrasov1994,Giustino2007},
an efficient method  to get estimates for the 
scattering times and mobilities relies on
deformation-potential  theory\cite{Bardeen:1950} and Fermi's golden
rule to estimate the scattering times. An estimate of the 2D mobility can be then obtained
using Takagi's formula~\cite{Takagi,Xi:2012,Qiao2014}
\begin{equation}
\mu_{\rm 2D} = \frac{e\hbar^3C_{\rm 2D}}{k_{\rm B}T m^*_{\rm e}  m_{\text{DOS}}(E^i)^2},
\end{equation}
where $m_{\rm e}^*$ is the transport effective mass, $m_{\text{DOS}}$ the
DOS effective mass (Table 1), $k_\text{B}$ the Boltzmann constant, $T$ the temperature,
$E^i$ the deformation potential constant, and $C_{\rm 2D}$ the elastic
modulus.  In our case, we need to consider this formula in the
multi-valley, anisotropic case: details can be found in Supplementary 
Notes 3--9 and described in Supplementary Figures 3--7, as well as the values of the extracted relevant
parameters (deformation potentials and elastic moduli, reported in Supplementary Tables 1 and 2, respectively). 

We would like to emphasise, however, that this formula, while often
adopted in the literature, cannot be used to obtain a quantitative estimate of
the mobility. Indeed, the formula neglects the coupling with ZA phonons (which may 
be important, as already discussed above), as well as with TA and optical phonons. Moreover, it cannot fully
capture the anisotropy of the electron-phonon coupling coefficients. 
A full ab-initio treatment of the electron-phonon scattering is required, if a quantitative estimation is needed (see e.g.\@ discussions in Refs.~\onlinecite{Fischetti2016,Liao2015}).
Nevertheless, we provide here an estimate of what we will call hereafter Takagi's mobilities, mainly to allow to compare As and Sb with other 2D materials already investigated in the literature within the same level of theory. 
We have estimated that, in the worst case scenario, the values of
actual mobilities could be reduced
up to a factor of 8 when a full treatment of the electron-phonon coupling is adopted, including intervalley scattering.

The resulting
values of the electron Takagi's mobility $\mu_{\rm c}$ 
and the hole Takagi's mobility $\mu_{\rm h}$
are 635 and 1700~$\text{cm}^2\text{V}^{-1}\text{s}^{-1}$, respectively,
for As and 630 and 1737~$\text{cm}^2\text{V}^{-1}\text{s}^{-1}$ for Sb.

The values of the electron Takagi's mobility are quite promising
and comparable with theoretical results for phosphorene using the same
level of theory\cite{Qiao2014} and even better than MoS$_2$\cite{Cai2014} owing to the smaller deformation potential. The hole Takagi's mobility is even larger, and 
in particular much larger than the experimentally measured value of the mobility in
MoS$_2$ at room temperature\cite{Radisavljevic2013} and in other 2D materials, 
like e.g. phosphorene\cite{Liu2014}.
On the other hand, we note that our predicted values are smaller than 
those predicted by simulations at the same 
level of theory (Takagi's formula) for phosphorene\cite{Qiao2014}, owing to the larger elastic modulus and smaller deformation potential in the zigzag direction.

We also emphasise that in arsenene and antimonene the SOC effects are
negligible for the conduction band. Instead, $\mu_{\rm h}$ is significantly enhanced by the
SOC, due to the splitting of the topmost valence band and the resulting changes in the
effective masses and deformation potentials (see Supplementary Tables 1 to 3). In particular, the inclusion of the SOC increases $\mu_{\rm h}$ by 25\% and 84\% in As and
Sb, respectively (see Supplementary Table 4).

From these values of the Takagi's mobilities and the 
associated scattering times and carrier velocities reported
in the Supplementary Note 10, we estimate that the
mean free path limited by LA phonons is of the order of tens of nm. 
For this reason we assume
that, for the dimensions considered in this work, the use of the ballistic approximation is justified and sets a higher limit to the performance achievable
in these devices.

\begin{table}[tb]
\caption{\label{tab:effmasses}\textbf{Valence and conduction effective masses of arsenene and antimonene:} Effective masses of the relevant bands
  of arsenene and antimonene, in units of the electron mass $m_0$, when 
  SOC effects are included.
  Symbols are explained in the main text.  Note that
  $m^{\rm c}_{\text{DOS}}$ is the effective DOS mass for each of the
  6 identical conduction band valleys.}
\footnotesize
\begin{tabular}{lcc}
 & As & Sb \\\hline\hline $m^{\rm c}_{\text{DOS}}$ & 0.270 & 0.261
  \\ $m^{\rm c}_{\rm L}$ & 0.501 & 0.472 \\ 
  $m^{\rm c}_{\rm T}$ & 0.146 & 0.144 \\
  $m^{\rm  v}$ & 0.128 & 0.103 \\ 
\end{tabular}
\end{table}

\subsection*{Performance of arsenene and antimonene
 based devices}
\begin{figure} [tbp]
\begin{center}
\includegraphics[width=8cm]{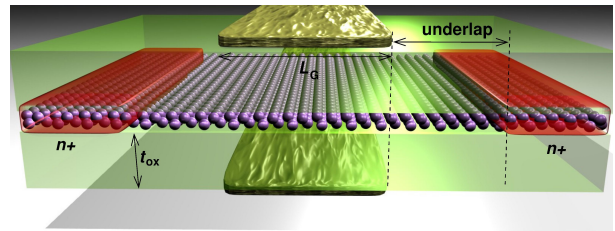}
\end{center}
\caption{\textbf{Double-gate n-doped MOSFETs:} Schematic view of the
double-gate n-doped MOSFETs studied here, where the channel is either 
an As or an Sb monolayer. In the figure, the doped contacts, the gate and 
the oxide are shown, together with the main geometrical parameters of the
device.}
\label{device}
\end{figure}

In view of the results of the previous section, we perform a full
device simulation of field-effect transistors based on
arsenene and antimonene as channel materials; we will focus
in particular on n-type devices since, as shown in
Supplementary Note 10, they show better performance as
compared to p-type FETs.
Our aim
is to assess quantitatively whether such devices can comply with
industry requirements for high-performance applications as needed by
the ITRS~\cite{ITRS}, which sets electrical and geometrical device
parameters to keep the pace with Moore's Law~\cite{Moore}.
The simulated device structure is shown in Fig.~\ref{device}. We consider
a double-gate transistor with doped source and drain, SiO$_2$ as
gate dielectric, and gate lengths ranging from 5 to 7~nm.  
The supply voltage ($V_{\rm DD}$) and the oxide thickness ($t_{\rm ox}$)
are chosen according to the device
channel length ($L_{\rm G}$), as specified by ITRS.
Spin-orbit coupling has been taken into account.

\begin{figure} [tbp]
\begin{center}
\includegraphics[width=8cm]{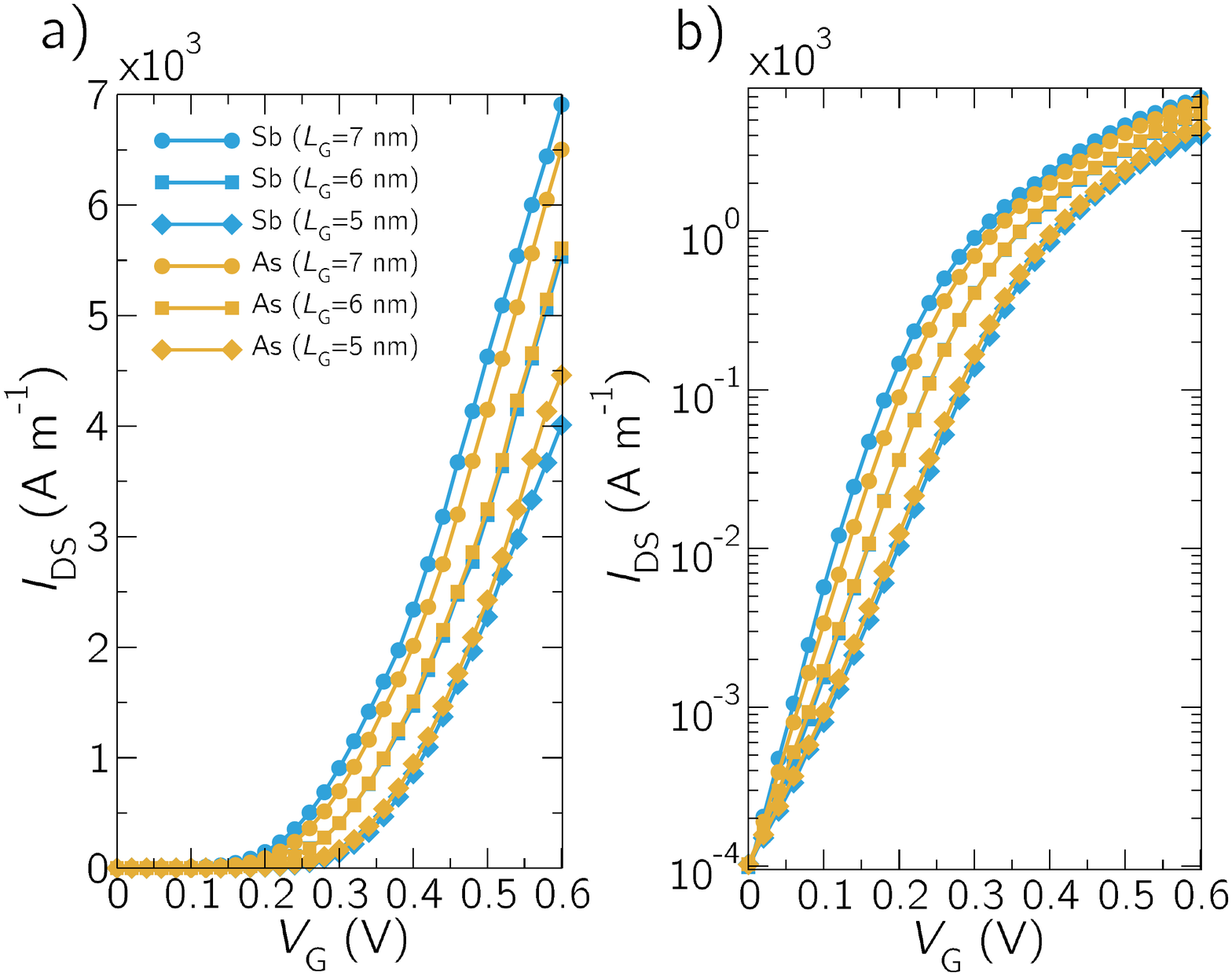}
\end{center}
\caption{\textbf{Transfer characteristics of As- and Sb-based MOSFETs:} $I_{\rm DS}-V_{\rm GS}$ curve in (a) linear and (b) semi-logarithmic 
  scale for Sb (light-blue lines) and As (yellow lines)
  transistors with $L_{\rm G}=7$~nm, $V_{\rm DS}=0.6$~V and $t_{\rm ox}=0.5$~nm
  (circles), $L_{\rm G}=6$~nm, $V_{\rm DS}=0.57$~V and $t_{\rm ox}=0.45$~nm
  (squares) and $L_{\rm G}=5$~nm, $V_{\rm DS}=0.54$~V and $t_{\rm ox}=0.42$~nm
  (diamonds).}
\label{IdVg}
\end{figure}

Figure~\ref{IdVg} shows the transfer characteristics of As- and 
Sb-based MOSFETs for the set of parameters listed in Table~\ref{tab1}.
For a fair comparison, the gate work function of all devices has
been shifted in order to have the same off-current $I_{\rm OFF}= 0.1$
A~m$^{-1}$ at $V_{\rm GS}=V_{\rm OFF}=0$~V, i.e., the smallest current driven by the transistor.
Both As and Sb transistors show similar $I-V$ characteristics as 
a consequence of their very similar conduction bands~\cite{Shengli2015}. 

From the $I-V$ characteristics, we can extract
the main figures of merit (FOM) required to assess the
device performance, that we summarise in Table~\ref{tab1}. 
In particular, the
subthreshold swing (SS), defined as the inverse slope of the
$I_{\rm DS}-V_{\rm GS}$ curve in semi-logarithmic scale in the subthreshold
regime, provides relevant information regarding the sensitiveness of
the device to short-channel effects:
the smallest SS achievable in thermionic devices at room temperature
is equal to 60 mV~dec$^{-1}$.~\cite{Taur}  For a gate length of 7 nm, both As and Sb
based MOSFETs exhibit excellent SS: 64~mV~dec$^{-1}$ and 60~mV~dec$^{-1}$,
respectively.  As the channel length gets shorter ($L_{\rm G}=6$
and 5~nm), SS increases to 81~mV~dec$^{-1}$ and 106~mV~dec$^{-1}$ for As, 
and 83~mV~dec$^{-1}$ and 106~mV~dec$^{-1}$ for Sb transistors, respectively. The reported values of SS for both materials show very promising performances, maintaining a subthreshold slope of approximately 100~mV~dec$^{-1}$ even for the smallest devices.

Another FOM is the $I_{\rm ON}$, i.e., the largest current driven by the transistor (for $V_{\rm GS}$=$V_{\rm DS}$=$V_{\rm DD}$.)
All our considered devices comply with $I_{\rm ON}$ requirements from ITRS.
It is important to say that in our calculation
the contact resistance has been neglected, and therefore our results
represent an upper limit for the achievable $I_{\rm ON}$.

The intrinsic delay time $\tau$ and the dynamic power indicator
(DPI) provide instead information regarding the switching speed and
the power consumption of a device, respectively.  The
values we obtain comply with ITRS requirements, even for
the shortest gate length.
In particular, DPI and $\tau$ are the energy and the time
it takes to switch a CMOS NOT port from the logic 1 to the logic 0 and viceversa, respectively.
In the same Table~\ref{tab1}, we also show the cut-off
frequency $f_{\rm T}$, i.e., the frequency for which the current gain
of the transistor is equal to one, which is a relevant parameter for radio-frequency
applications.
Both As and Sb MOSFETs exhibit excellent $f_{\rm T}$ compared to ITRS.
As compared to other two-dimensional materials, As and Sb show performance comparable
to that obtained in black phosphorus FETs~\cite{Lam2014,Liu2014bis}.

\begin{table*}
\caption{\label{tab1} 
\textbf{Performance of arsenene and antimonene n-MOSFETs:} 
  Device parameters and calculated figures of merit 
  of As- and Sb-based n-MOSFETs for
  different channel lengths. The target FOMs set by ITRS for end-of-the-roadmap
  are also indicated~\cite{ITRS}. The meaning of each parameter is explained
in the main text.}
\footnotesize
\begin{tabular}{lcp{0.5cm}lllp{0.5cm}lll}
 &&& As & & && Sb & & \\ \hline\hline
 $L_{\rm G}$ (nm) &&& 7 & 6 & 5 && 7 & 6 & 5
 \\ 
$t_{\rm ox}$ (nm) &&& 0.5 & 0.45 & 0.42 && 0.5 & 0.45 & 0.42
 \\ 
$V_{\rm DD}$ (V) &&& 0.6 & 0.57 & 0.54 && 0.6 & 0.57 & 0.54
\\ \hline\hline
SS
 (mV~dec$^{-1}$) &[this work]&& 64 & 81 & 106 && 60 & 83 & 106
 \\ \hline 
\multirow{2}{*}{$I_{\rm ON}$ (A~m$^{-1}$)} &[ITRS]&& $\ge 2.19\times 10^{3}$ &
 $\ge 2.31 \times 10^{3}$ & $\ge 2.41 \times 10^{3}$ && $\ge 2.19 \times
 10^{3}$ & $\ge 2.31 \times 10^{3}$ & $\ge 2.41 \times 10^{3}$
 \\ 
  &[this work]&& $6.57 \times 10^{3}$ & $4.9
 \times 10^{3}$ & $3.2 \times 10^{3}$ && $6.91 \times 10^{3}$
 & $4.93 \times 10^{3}$ & $2.98 \times 10^{3}$ \\ \hline
\multirow{2}{*}{$\tau$ (ps)} &[ITRS]&
 & $\le 0.125$ & $\le 0.1$ & $\le 0.08$ && $\le 0.125$ & $\le 0.1$ & $\le 0.08$ \\ 
 &[this work]&&
 0.04 & 0.045 & 0.052 && 0.042 & 0.047 & 0.055 \\
\hline \multirow{2}{*}{$f_{\rm T}$ (THz)} &
 [ITRS] && $\ge 1.91$ & $\ge 2.36$ & $\ge 2.88$ && $\ge 1.91$ & $\ge 2.36$ & $\ge 2.88$ \\ 
 &[this work]&& 5.8 & 6.01 & 5.47 && 5.51 & 5.94 & 4.83 \\ \hline
\multirow{2}{*}{DPI (10$^{-10}$ J~m$^{-1}$)} &[ITRS] && $\le 1.6$ & $\le 1.4$ & $\le 1.2$ && $\le 1.6$ & $\le 1.4$ & $\le 1.2$ \\ 
 &[this work]&& 1.58 & 1.25 & 0.91 && 1.76 & 1.31 & 0.89
\end{tabular}
\end{table*}

\begin{figure} [tbp]
\begin{center}
\includegraphics[width=8cm]{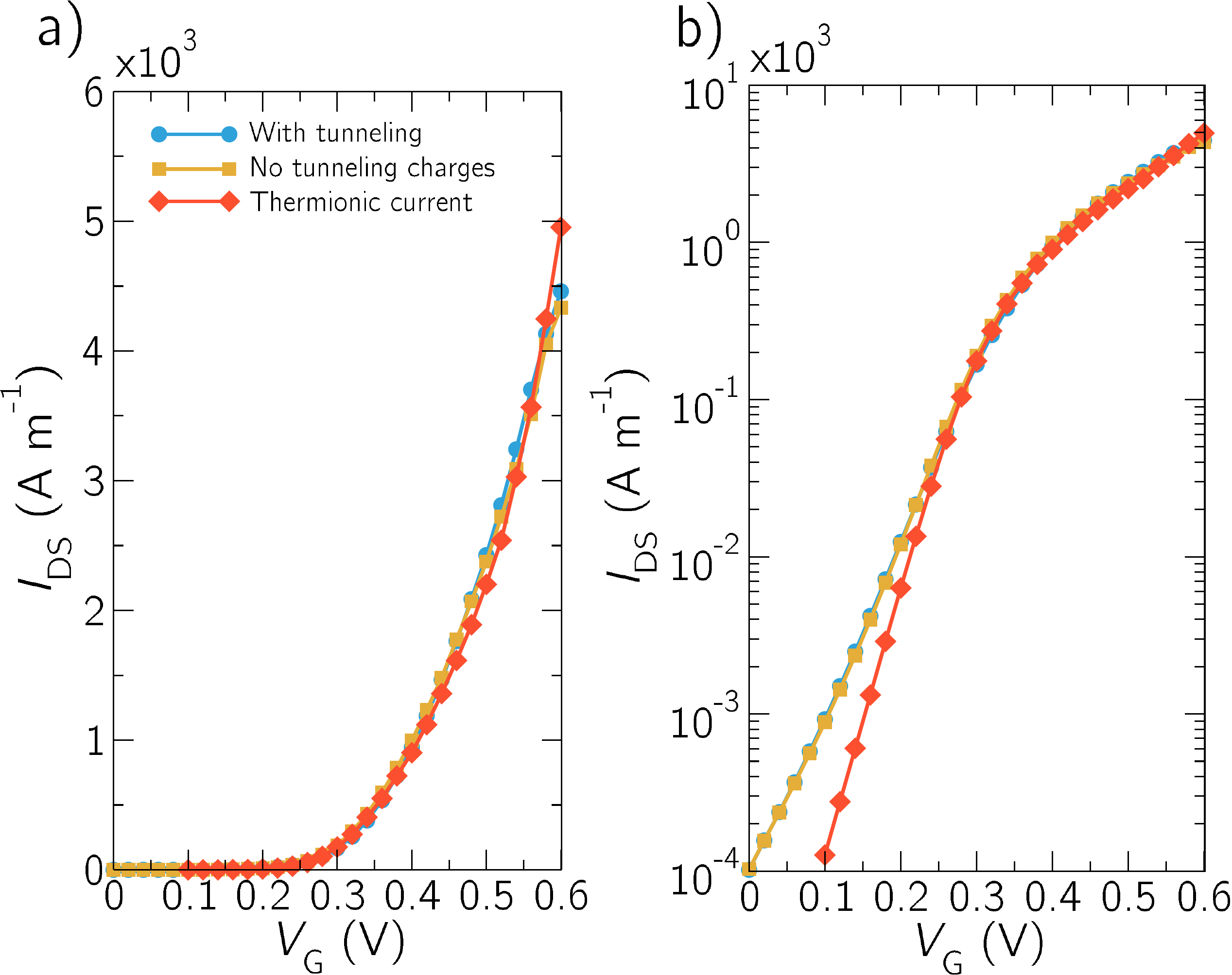}
\end{center}
\caption{\textbf{Short-channel effects on sub-threshold swing:} $I_{\rm DS}-V_{\rm GS}$ curve in (a) linear and (b)
  semi-logarithmic scale for As transistors with $L_G=5$~nm and
  $V_{\rm DS}=0.54$~V. The full simulation model considering tunnelling both in
  the current and the charges is represented by light-blue circles,
  whereas suppressed charges in the channel and thermionic currents only
  are represented by yellow squares and red diamonds, respectively.}
\label{Ftunnel}
\end{figure}

To get a deeper understanding of the effects limiting 
the device performance, we focus in particular on the degraded
subthreshold swing observed in short-channel devices,
which can be attributed to
two main phenomena: large tunnelling currents through 
the narrow barrier; and large parasitic capacitance at
source/drain-channel junctions, i.e., short channel effects.  To
elucidate which of the two effects plays a major role, 
we consider them separately for the 
shortest device: we either neglect quantum phenomena for
the current (i.e., tunnelling through the channel barrier), but not
when computing the charge (i.e., we consider mid-gap tunnelling states
when solving the electrostatic problem, red line in
Fig.~\ref{Ftunnel}) or viceversa (yellow line in Fig.~\ref{Ftunnel}).
Transfer characteristics with almost ideal SS ($\sim
60$~mV~dec$^{-1}$) are obtained in the first case, demonstrating that the SS
in short-channel devices is limited by the fact that the channel
barrier is almost transparent for electrons injected from the source
reservoir, and not by the short-channel effects, as one may expect
for such short channel lengths.
This suggests that, from an engineering point of view, in order to
improve the performance in ultra-scaled devices, efforts have to be
directed in increasing the opacity of the channel barrier. This
can be achieved for example exploiting materials with larger 
longitudinal tunnelling effective masses, or using uniaxial strain to split
the conduction valleys while selecting bands with
larger tunnelling effective mass in the transport direction.

\begin{figure} [tbp]
\begin{center}
\includegraphics[width=8cm]{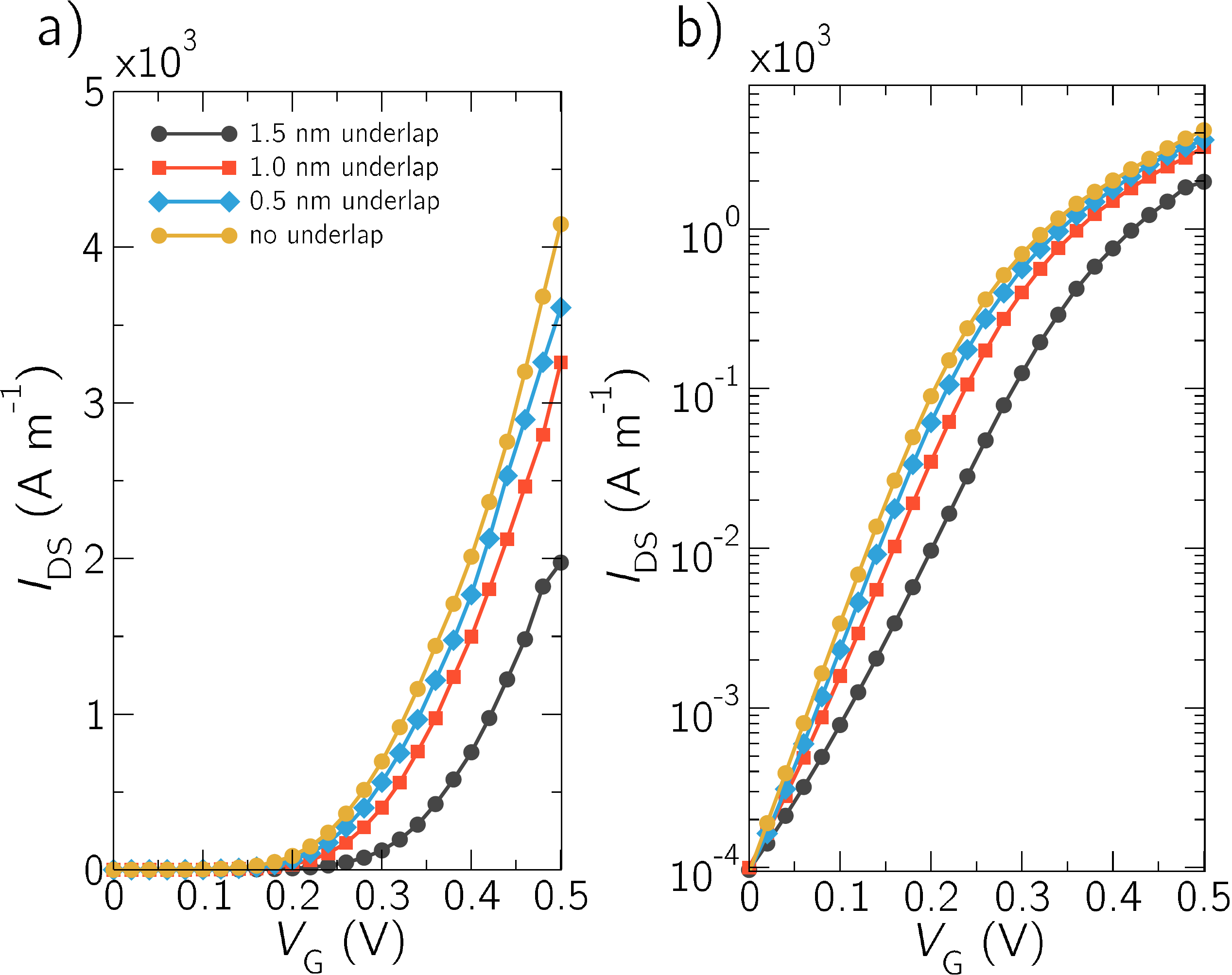}
\end{center}
\caption{\textbf{Transfer characteristics for different underlap:} $I_{\rm DS}-V_{\rm GS}$ curve in (a) linear and (b)
  semi-logarithmic scale for As transistors with total distance between
  drain and source electrodes
  of $7$~nm, and different underlap.}
\label{underlapnew}
\end{figure}

\begin{figure} [tbp]
\begin{center}
\includegraphics[width=8cm]{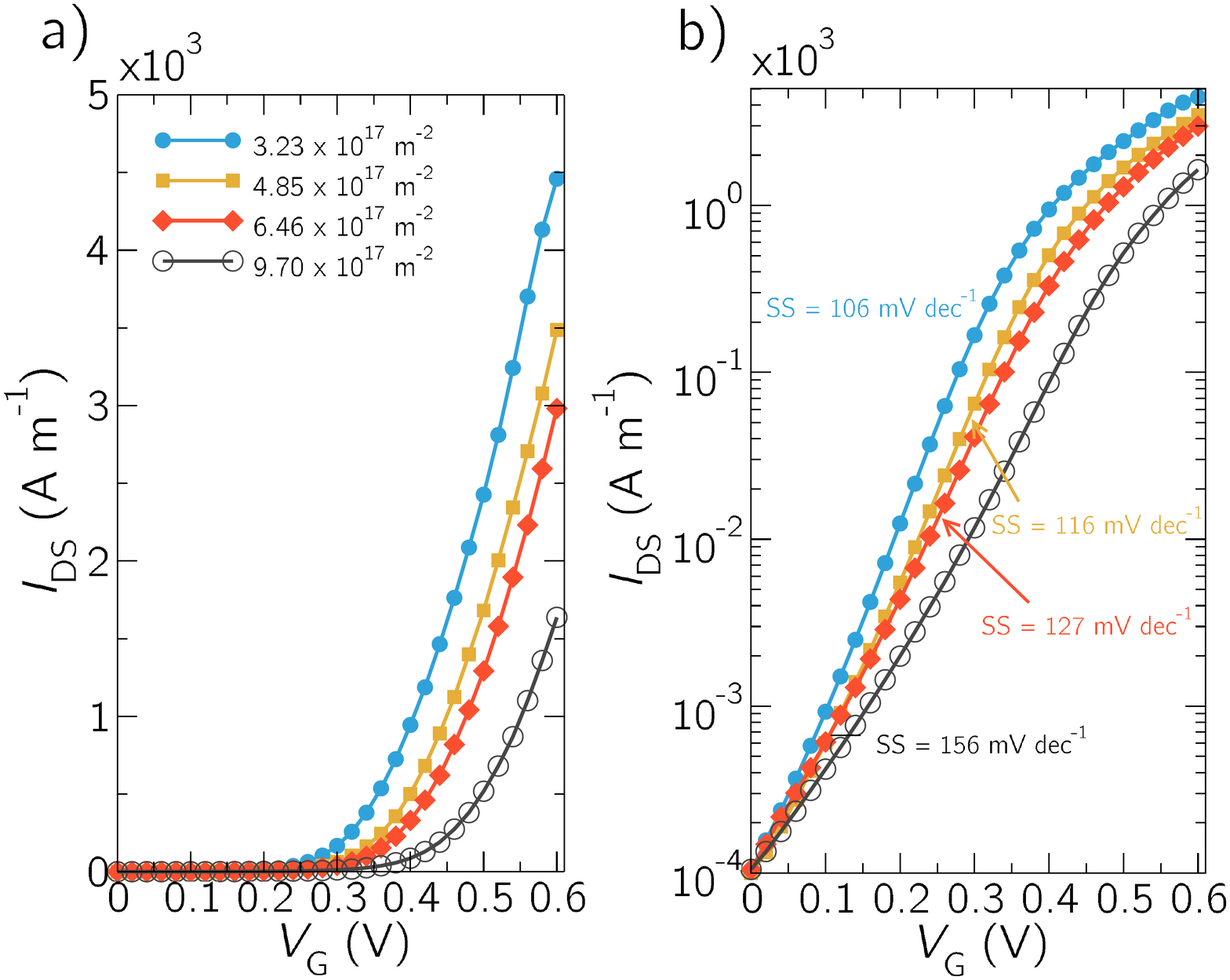}
\end{center}
\caption{\textbf{Boosting device performances by tuning source and drain doping concentrations:} $I_{\rm DS}-V_{\rm GS}$ curve in (a) linear and (b)
  semi-logarithmic scale for As transistors with $L_{\rm G}$ = 5~nm and for
  different source and drain doping concentrations.}
\label{doping}
\end{figure}

Performing an investigation along the device parameter space, we have 
also computed the $I-V$ characteristics for different gate underlap values (defined
in Fig.~\ref{device}), fixing the distance between source
and drain electrodes (i.e., 7~nm) and changing accordingly the gate length $L_{\rm G}$ 
and the underlap region, and the source and drain doping concentrations
(Fig.~\ref{underlapnew} and Fig.~\ref{doping}, respectively).
As it can be seen from the results reported in Fig.~\ref{underlapnew}, 
$I-V$ curves change only marginally when considering different underlap values.
As a consequence, from a fabrication point of
view, while control of the geometrical parameters for the gate
contacts is required, minor dispersions do not drastically degrade
the device performance. From Fig.~\ref{doping}, instead,
we deduce that the sub-threshold slope improves significantly when the doping is reduced.
Therefore, the doping concentration can be used as an additional parameter to 
optimise the device performance.

\section*{Discussion}
In summary, we have provided a comprehensive analysis of 2D FET transistors based on
arsenene and antimonene (i.e., monolayers composed of As and Sb, respectively),
demonstrating that these materials are promising for 
high-performance devices for digital applications. 
Our single-valley and multi-valley upper estimates of the mobilities in the Takagi's approximation show that high phonon-limited mobilities 
can potentially be obtained both in monolayer As and Sb, even if ab-initio simulations of the electron-phonon scattering are required to obtain quantitative predictions for this quantity.
This result motivated our extensive investigation of
device performance using a multiscale approach,
where the predictive power of density-functional-theory calculations 
has been incorporated 
into an efficient tight-binding model using maximally-localised Wannier functions.
When exploited as channel materials in field-effect transistors, arsenene
and antimonene show a performance compliant with industry requirements for 
ultra-scaled channel lengths below 10~nm, where the ultimate atomic thickness of
the exploited 2D materials effectively manages to
suppress short-channel effects and tunnelling starts to play a major role.
We expect therefore that our predictions will provide a strong motivation for 
further experimental investigation of these novel materials.

\section*{Methods}
\textbf{First-principles calculations} -- All first-principles
calculations have been performed using density-functional theory (DFT)
as implemented in the pw.x code of the Quantum ESPRESSO v.5.1.2
distribution~\cite{Giannozzi2009}, using the PBE
exchange--correlation functional~\cite{Perdew1996}. Pseudopotentials and energy cutoffs
in a plane-wave basis have been chosen using the converged results
provided in the SSSP pseudopotential library~\cite{sssp} 
for calculations without SOC. In
particular, we used an ultrasoft pseudopotential~\cite{Vanderbilt1990} from
PSLibrary~\cite{DalCorso2014} with cutoffs of 40 and 320~Ry (for the
expansion of the wavefunctions and the charge density, respectively)
for As; and an ultrasoft pseudopotential from the GBRV library~\cite{Garrity2014}
(with cutoffs of 50 and 400~Ry, respectively) for Sb.  
For calculations including SOC we used instead 
norm-conserving pseudopotentials from the Pseudo Dojo project~\cite{pseudodojo} 
with cutoffs of 40 and 160~Ry (for the
expansion of the wavefunctions and the charge density, respectively)
for As; and with cutoffs of 80 and 320~Ry, respectively, for Sb.
Supercells with
20~\AA{} of vacuum in the direction orthogonal to the layers have been
considered to minimise the interaction between periodic
replicas. Integrals on the Brillouin zone have been performed on a
$14\times 14\times 1$ $\Gamma-$centred grid for the primitive cell
(two atoms) and on a $6\times 10\times 1$ $\Gamma-$centred grid for
the rectangular supercell (containing four atoms). Convergence on the
charge density in the self-consistent loop was considered achieved
when the estimated energy error was smaller than $1\times
10^{-8}$~meV. Structures have been relaxed using the
Broyden--Fletcher--Goldfarb--Shanno algorithm until forces were
smaller than 1~meV~\AA$^{-1}$.  The same parameters have been used also for
relaxing atomic positions at fixed cell for the evaluation of elastic
moduli and deformation potentials.

\textbf{Wannier functions} -- MLWF~\cite{Marzari1997,Souza2001,Marzari2012} have
been computed using v.2.1 of the Wannier90
code~\cite{Mostofi2008,Mostofi2014}.  The same k-grids used for the
computation of the DFT charge densities have been employed to compute the
wavefunctions and overlap matrices used as input to calculate Wannier
functions. The lower-energy bands have been explicitly excluded from
the calculation: with the pseudopotentials we used, 2 for As and 12 for
Sb in the primitive cell (4 and 24, respectively, 
when considering explicitly the spin degeneracy), 
and only 6 bands (12 with spin degeneracy) around the fundamental gap
have been considered. We have chosen $p$-type orbitals centred on
each atom in the cell as initial projections. Convergence has been
considered achieved when the change in the total spread was smaller
than $10^{-12}$~\AA$^2$ for at least 20 iterations. Wannier functions
have then been used to compute band structures and density of states on
denser $\vec{k}-$grids ($400\times700$ in the rectangular cell).  DFT and
Wannier calculations have been managed using the AiiDA
framework\cite{Pizzi2016} v. 0.5.0 to manage, automate and store in a
graph database calculations, results, and computational workflows
(e.g., for band structure calculations, Wannierisation, effective mass
evaluations).

\textbf{Device simulations} -- The Hamiltonian expressed on the MLWF basis set has been exploited in order
to compute transport within the NEGF formalism~\cite{Datta}. 
The system is considered infinite along the zigzag direction (with Bloch periodic boundary conditions), while the transport channel is along the armchair direction.
To compute the currents in the device in the ballistic regime, 
we have used the open-source NanoTCAD ViDES~\cite{ViDES} code. 
In particular, in order to accurately
reproduce the energy bands obtained from first-principles, up to 58
nearest-neighbours have been included in the Hamiltonian, and transport
problems have been solved considering 30 wavevectors in the Brillouin
zone and an energy step of 1~meV.  For the electrostatic
problem, the two-dimensional Poisson equation has been solved, while
potential translational invariance has been considered in the
direction transversal to transport. All transport calculations are performed at room temperature. In all device simulations (except where explicitly otherwise mentioned) a doping concentration of $3.23 \times 10^{17}$ m$^{-2}$ has been considered for the source and drain contacts.

\section*{Acknowledgements}
We thank Nicolas Mounet for providing us with the AiiDA workflow to 
compute band structures with Quantum ESPRESSO and Massimo Fischetti for very insightful discussions. 
This work was supported by a grant from the Swiss National
Supercomputing Centre (CSCS) under project IDs s580.
M.G., N.M., G.I. and G.F. gratefully acknowledge the Graphene Flagship
(contract 604391).

\section*{Supplementary Information}

\setcounter{equation}{0}
\renewcommand\theequation{S\arabic{equation}}

\stepcounter{myseccnt}

\renewcommand\thefigure{\arabic{figure}}
\renewcommand\thetable{\arabic{table}}
\renewcommand\figurename{Supplementary Figure}
\renewcommand\tablename{Supplementary Table}
\setcounter{secnumdepth}{2} 
\renewcommand\thesubsection{\arabic{subsection}}
\titleformat{\subsection}{\small\bfseries
}{}{0em}{Supplementary Note \thesubsection{}: #1}[\vspace*{-0.5em}]%

\subsection{Band structure of antimonene, and spin-orbit coupling effects}
In the paper, only the band structure of arsenene has been shown, because the band
structure of antimonene is qualitatively the same. For completeness we report in this Supplementary Information, in Supplementary Figure~\ref{fig:sbbands}, the band structure of antimonene as well as the bands interpolated using maximally-localised Wannier functions. Moreover, we compare all results with the case in which the spin-orbit coupling (SOC) has been set to zero, showing that the SOC almost does not change the final results for the mobilities in the conduction band, while it significantly affects results in the valence (due to the splitting of the degeneracies). In particular, in Supplementary Figure~\ref{fig:soc-nosoc-bands} we compare the (Wannier-interpolated) band structures of As and Sb, to show the effect of the spin-orbit coupling on the electronic bands.

\begin{figure}[tb]
  \centering
  \includegraphics[width=0.9\linewidth]{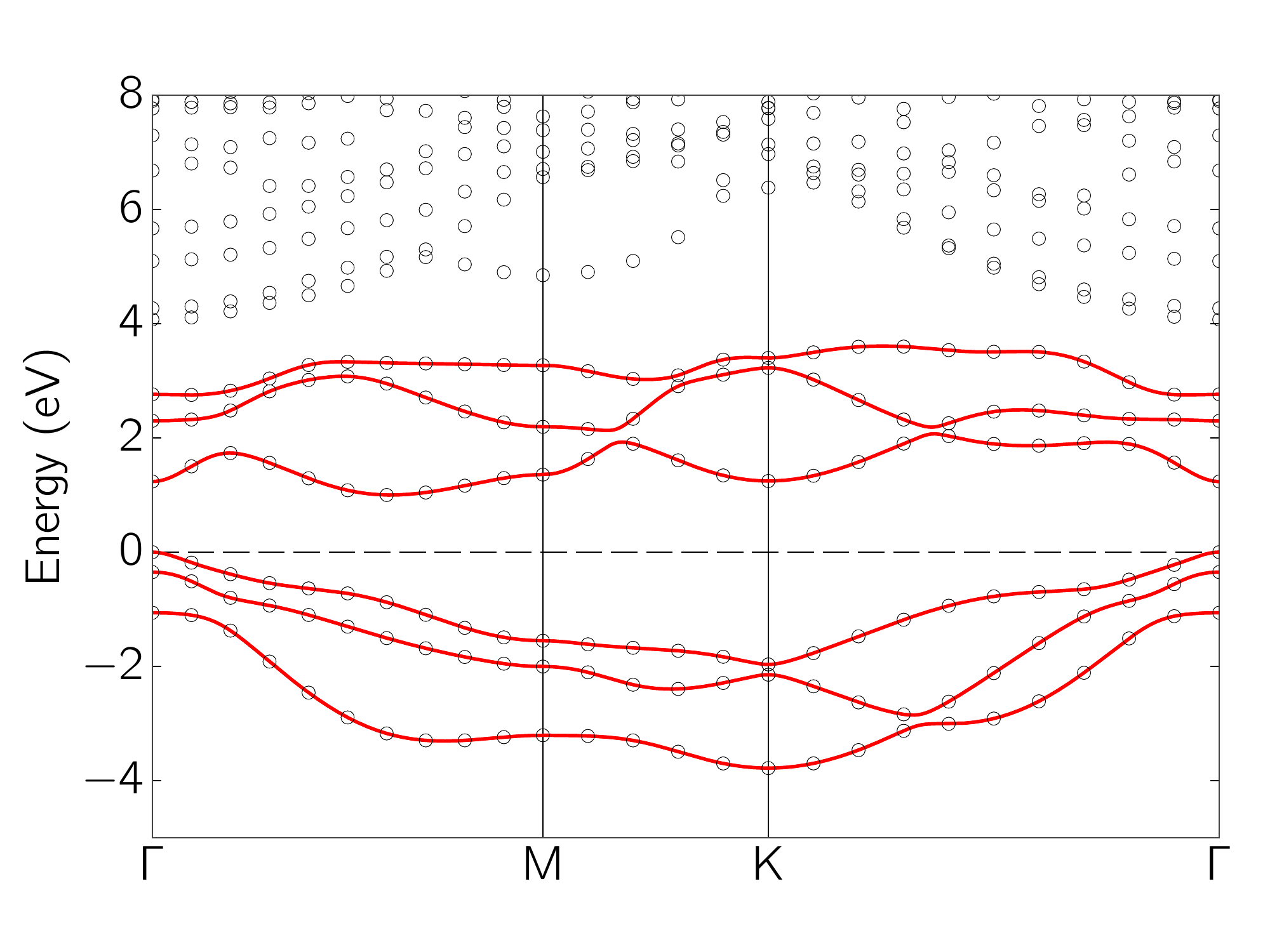}
  \caption{Energy bands of antimonene along a
high-symmetry path in the Brillouin zone, including spin-orbit coupling effects. Empty circles denote the results
of a direct DFT calculation while red solid lines represent the
Wannier-interpolated bands. The zero of energy is set to the top of the valence bands.\label{fig:sbbands}}
\end{figure}

\begin{figure}[tb]
  \centering
  \includegraphics[width=\linewidth]{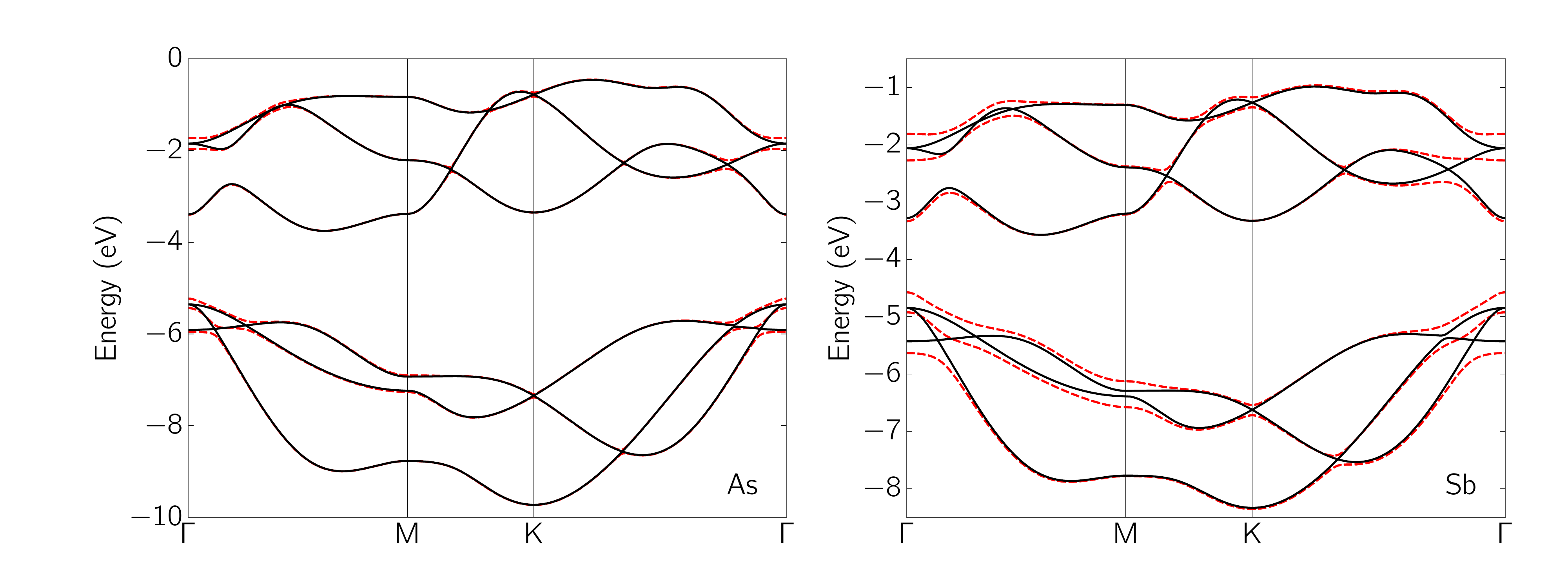}
  \caption{Wannier-interpolated energy bands of arsenene (left panel) and antimonene (right panel) along a high-symmetry path in the Brillouin zone, comparing the band structure calculated including (red dashed lines) and disregarding (solid black lines) spin-orbit coupling effects. The bandgaps without SOC are 1.62~eV and 1.29~eV for As and Sb, respectively, while the valence-band splitting due to the SOC is 0.208~eV and 0.349~eV, respectively. The values of the bandgaps with SOC are reported in the main text. The zero of energy is set at the vacuum level.\label{fig:soc-nosoc-bands}}
\end{figure}

\subsection{Calculation of mobilities}
We describe in detail in this Section how we calculated the phonon-limited
mobilities of the 2D systems considered in the paper.

We calculate mobilities using the Boltzmann transport equation, where
the relaxation time for scattering with LA phonons is computed using deformation
potential theory~\cite{Bardeen:1950} in the effective mass
approximation. For a single-valley, non-degenerate band, the reciprocal
of the scattering time with LA phonons propagating in the $\beta$
direction for the $i-$th band at $\vec k-$point $\vec k$ 
is given by~\cite{Bardeen:1950,Xi:2012}:
\begin{equation}
\frac 1 {\tau_\beta(i,\vec k)} = \frac {2\pi k_{\text{B}} T (E_\beta^i)^2}{\hbar
  C_\beta} \sum_{\vec k'}\delta\left[\eps_i(\vec k) - \eps_i(\vec
  k')\right](1-\cos \theta),
\end{equation}
where $k_{\text{B}}$ is the Boltzmann constant, $E_\beta^i$ is the deformation
potential of the $i-$th band for deformations in the $\beta$
direction, $C_\beta$ is the 2D elastic modulus for strains along
$\beta$, $\eps_i(\vec k)$ is the energy of the $i-$th band at $\vec
k$, and we are replacing for simplicity the scattering angle
weighting factor with $(1-\cos\theta)$ (valid for a spherical energy
surface, where $\theta$ is the angle between $\vec k$ and $\vec k'$).

The calculation of the various coefficients is
described in the next sections. To get the value of the mobility, though,
it is easier to work out the formula in the specific
case of a 2D system. By replacing the sum over $\vec k'$ with an
integral, and then passing from an integral over the
Brillouin Zone to an integral over energies, we obtain that
\begin{equation}
\label{eq:oneovertau}
\frac 1 {\tau_\beta} = \frac{k_\text{B} T (E_\beta^i)^2 (m_{\text{DOS}}^*)_i}{\hbar^3C_\beta}.
\end{equation}
Finally, the mobility $\mu$  is diagonal relative to the axes of the
effective mass of the valley of interest~\cite{Herring:1956}, 
and its diagonal components can be obtained as
\begin{equation}
\label{eq:mobilityoneband}
\mu_{\beta\beta} = e\braket{\tau_\beta}\cdot\left(\frac{1}{m^*}\right)_{\beta\beta}
\end{equation}
with 
$e$ being the electron charge,  $(1/m^*)$ the inverse effective mass tensor,
and $\braket{\tau}$ the
average scattering time (as defined in Eq.~(16) of Ref.~\cite{Herring:1956}).
The formula is valid in the same reference frame in which the effective mass
tensor is diagonal.
Note that in the 2D case, $\braket{\tau} = \tau$ since there is no energy 
dependence.

In the case of multiple valleys or degenerate bands, the expression
for the mobility is the same of Supplementary Equation~\eqref{eq:mobilityoneband}, but
applies to a single valley. To understand how the mobilities need to
be added, it is easier to write them in terms of the electrical
conductivity $\sigma$, using the fact that the total conductivity is the
sum of the conductivities of the different channels. 
To do so, we notice that we can write the mobility $\mu^i$ of 
Supplementary Equation~\eqref{eq:mobilityoneband} as
\begin{equation}
\mu^i = \frac {\sigma^i}{n^i e},
\end{equation}
for each valley $i$ contributing to the transport, where 
$n^i$ is the charge in the $i-$th valley.
Using $\sigma_{\text{T}}= \sum_i \sigma^i$, where $\sigma_{\text{T}}$ represents the total
conductivity, we obtain therefore for the total mobility:
\begin{equation}
\label{eq:mobilitytotal}
\mu_\text{T} = \frac{\sum_i \mu^i n^i}{\sum_i n^i}.
\end{equation}

\subsection{Relevant band edges: generic discussion}
As already discussed in the paper, the bands of As and Sb are quite
similar, so the following discussion applies to both systems, unless
explicitly mentioned.

As it is visible from the band structure of the two materials (see
Supplementary Figure~\ref{fig:sbbands}, and Figure~1 in the main paper), 
the relevant band edge to consider in the conduction is the minimum along
the $\Gamma-$M line. All other conduction band edges (for instance,
the minimum at $\Gamma$ or at the K point) are several hundreds of
meV above the one along $\Gamma-$M, and therefore do not contribute to
transport at room temperature for typical doping levels. 
We have
also verified that this condition holds at least for a strain range
between $-1$ and $1\%$. In the unstrained case, this minimum is 
composed of six degenerate valleys, along the six equivalent 
$\Gamma-$M lines. In the
rectangular supercell shown in blue in Supplementary Figure~\ref{fig:realspaceandBZcell-2xsupercell}(a), two of these lines fold along
the $\Gamma-$X line, while four fold on the $\Gamma-$S lines, as shown
in Supplementary Figure~\ref{fig:realspaceandBZcell-2xsupercell}. When a
uniaxial strain is applied, the degeneracy is lost and 
the six valleys split in two groups 
(the two $\Gamma-$X and the four $\Gamma-$S valleys), with
different deformation potentials. We also stress that in the rectangular cell,
the $\Gamma-$S line is not anymore a high-symmetry line, so the band
edge can also move out of this line.
Note that, actually, for the evaluation of the mobility 
we just need the deformation potential along the two
principal axes of the effective mass, that is, we just need the values
for the $\Gamma-$X lines, both for strains along the armchair and zigzag
directions. This discussion holds both with and without SOC.

In the valence, in the case without SOC, 
the only relevant band edges are the two degenerate
maxima at $\Gamma$ (also in this case other local maxima are further
down in energy and can be disregarded for the calculation of the
mobility). These are both single-valley maxima, but they split
when a uniaxial strain is applied because they have different
deformation potentials. 
When including SOC effects, the degeneracy is split and, for realistic
band filling levels ($\lesssim 10^{13} \text{ cm}^{-2}$), we can limit our
calculations only to the topmost valence band, since it is the only ones
going to be filled.

Let us now compute an explicit formula for the mobilities for the As and Sb 
monolayer systems:

\subsection{Relevant conduction band edges}

\begin{figure}[tb]
  \centering
  \includegraphics[width=0.7\linewidth]{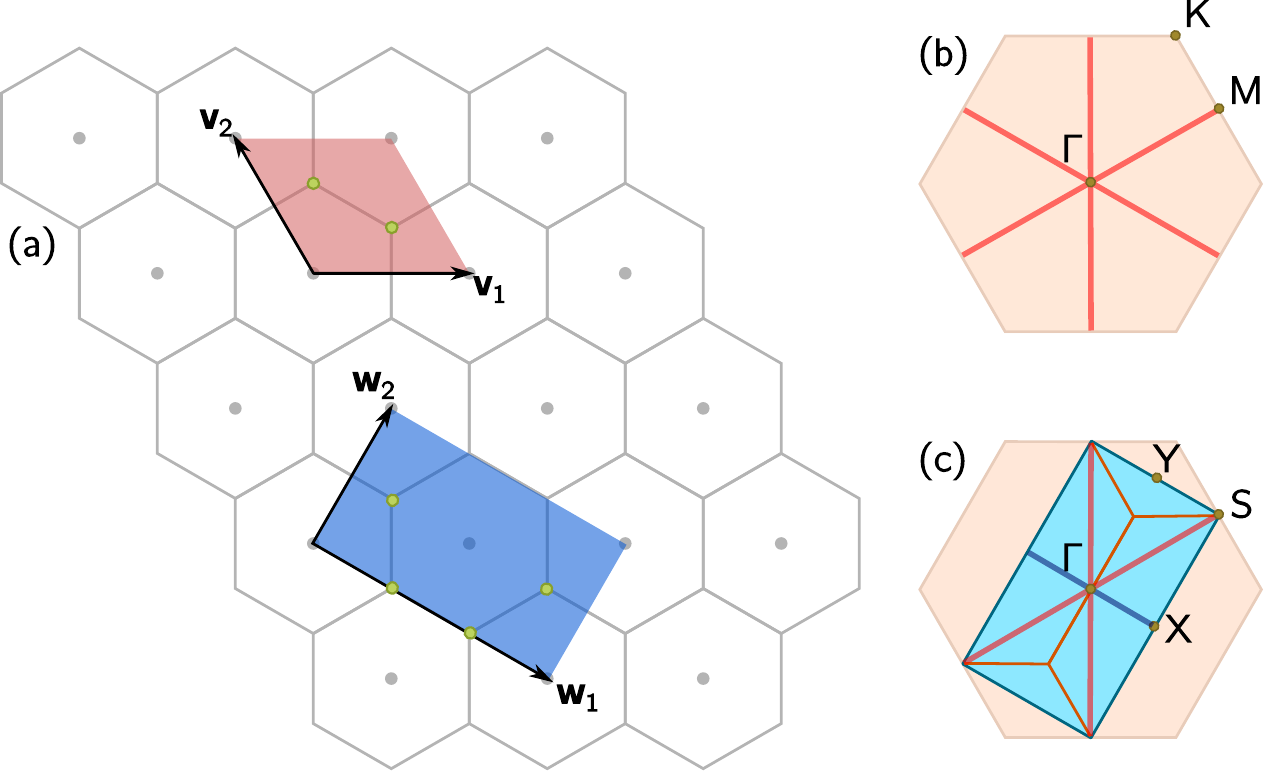}
  \caption{(a) Primitive unit cell with two atoms (top), 
  and rectangular supercell 
  with four atoms (bottom) used when applying strain in a given direction $\beta$. 
  In particular, $\vec w_1$ ($\vec w_2$) is along an armchair (zigzag) direction. Both
  $\vec v_1$ and $\vec v_2$ are along zigzag directions, instead. (b) Brillouin zone 
  of the primitive unit cell. The six equivalent $\Gamma-$M lines are highlighted
  in red. (c) Folding of the Brillouin zone when the 4-atom supercell is considered 
  (light blue rotated rectangle), 
  and labelling of the high-symmetry points. The six $\Gamma-$M lines of panel (b)
  split into four $\Gamma-$S lines (red) and two $\Gamma-$X lines (dark blue).
  \label{fig:realspaceandBZcell-2xsupercell}}
\end{figure}

The mass tensor of each of the 6 $\Gamma-$M valleys is non-isotropic, 
with a longitudinal mass $m^\text{c}_\text{L}$ along
  the $\Gamma-$M direction, and a transverse mass $m^\text{c}_\text{T}$ in the orthogonal 
direction. The inverse-mass tensor in a basis set where the first vector 
is along the longitudinal direction is therefore:
\begin{equation}
  \left(\frac 1 {m^*}\right)_{ij} = \left(\begin{array}{cc}
      \frac 1 {m^\text{c}_\text{L}} & 0 \\
      0 & \frac 1 {m^\text{c}_\text{T}} \\
    \end{array}
    \right).
\end{equation}
In this basis set, the two $\tau_\beta$ to calculate are along the longitudinal
direction (corresponding to strains along the armchair direction,
see Supplementary Figure~\ref{fig:realspaceandBZcell-2xsupercell}), and along the
transverse direction (zigzag). 
Therefore, substituting Supplementary Equation~\eqref{eq:oneovertau} into Supplementary Equation~\eqref{eq:mobilityoneband}, 
the mobility tensor for a single valley, in this basis set, will be:
\begin{equation}
\label{eq:mobility-singlevalley}
\boxed{\mu^{\text{cond},i} = \frac{e\hbar^3}{k_\text{B}T m_{\text{DOS}}^\text{c}}\left(\begin{array}{cc}
      \frac {C_{\text{armchair}}} {(E^{\text{cond}}_{\text{armchair}})^2 m^\text{c}_\text{L}} & 0 \\
      0 & \frac {C_{\text{zigzag}}} {(E^{\text{cond}}_{\text{zigzag}})^2 m^\text{c}_\text{T}} \\
    \end{array}
    \right),}
\end{equation}
with $m_{\text{DOS}}^\text{c}$ being the DOS mass of a single
  conduction valley.

Since the DOS effective mass is the same for each valley and the valleys are
degenerate (in the
  absence of strain that splits the bands, as discussed above),
all valleys have the same population $n^i=n/6$ and therefore
the total conductivity is simply:
\begin{equation}
\mu^{\text{cond}}_\text{T} = \frac 1 6 \sum_{i=0}^5 \mu^{\text{cond},i},
\end{equation}
where the tensors $\mu^{\text{cond},i}$ must, however, be rotated.
We remind here that given a tensor 
$\begin{pmatrix}\alpha & 0 \\0 & \beta\end{pmatrix}$ in a reference
frame in which it is diagonal, its form 
in a frame rotated counterclockwise by an angle $\theta$ is
\begin{equation}
\begin{pmatrix}
  \alpha\cos^2\theta + \beta\sin^2\theta & (\beta-\alpha)\cos\theta\sin\theta \\
 (\beta-\alpha)\cos\theta\sin\theta &   \alpha\sin^2\theta + \beta\cos^2\theta\end{pmatrix}.
\end{equation}

Choosing the reference frame in which $\mu^{\text{cond},0}$ is diagonal, 
we have to sum the 6 bands, each rotated by $\frac \pi 3$ with respect
to the previous one, and therefore
\begin{equation}
(\mu_\text{T}^{\text{cond}})_{11} = \frac\alpha 6 \sum_{i=0}^5 \cos^2\left(i\cdot \frac \pi 3\right) + 
\frac \beta 6 \sum_{i=0}^5 \sin^2\left(i\cdot \frac \pi 3\right).
\end{equation}
Now $\sum_{i=0}^5 \sin^2\left(i\cdot \frac \pi 3\right)= 0 + \frac 3 4 + \frac 3 4 +0 + \frac 3 4 + \frac 3 4 = 3$ and $\sum_{i=0}^5 \cos^2\left(i\cdot \frac \pi 3\right) = \sum_{i=0}^5 1-\sin^2\left(i\cdot \frac \pi 3\right) = 6-3=3$, and
therefore $(\mu_\text{T}^{\text{cond}})_{11} = (\mu_\text{T}^{\text{cond}})_{22} = \frac {\alpha+\beta}2$. Similarly, one can show that the off-diagonal contributions cancel in 
pairs, and therefore the total conduction mobility is a multiple of the identity, with value
\begin{equation}
\boxed{
\mu_\text{T}^{\text{cond}} =  \frac{e\hbar^3}{2k_\text{B}Tm_{\text{DOS}}^\text{c}}\left(
      \frac {C_{\text{armchair}}} {(E^{\text{cond}}_{\text{armchair}})^2 m^\text{c}_\text{L}} +
       \frac {C_{\text{zigzag}}} {(E^{\text{cond}}_{\text{zigzag}})^2 m^\text{c}_\text{T}} \right).
}
\end{equation}
The fact that the $\mu_\text{T}$ tensor is isotropic is expected, 
because the system has hexagonal symmetry. 
Note that averaging on all valleys is equivalent to 
averaging the tensor of a single valley in all directions, as it is the 
case also for cubic systems~\cite{Herring:1956}.

\subsection{Relevant valence band edges}

Without SOC, in the valence band we have instead two single
  valleys that are degenerate at $\Gamma$, but with different masses
  (and scattering times), that we can call light holes (LH) and heavy
  holes (HH). 
  We assume that both bands are parabolic, and in this case the 2D
  density of states is a step function (where the step height,
  occurring at the band edge energy, is proportional to the 2D DOS
  effective mass). Then, the population $n_i$ of a given band is simply
\begin{equation}
n_i = D\cdot m^i_{\text{DOS}}
\end{equation}
where $D$ is a constant that contains the energy difference between
the chemical potential and the band edge. (We are assuming that
this energy difference is the same for both bands, true if
intraband scattering events can quickly equilibrate the bands so that they
have the same chemical potential, and if we consider an unstrained
system so that the two LH  and HH bands are degenerate at their
maximum).

Replacing this simple expression for $n_i$ in 
Supplementary Equations~\eqref{eq:mobilityoneband} and~\eqref{eq:mobilitytotal}, the mobility is:
\begin{equation}\small
\mu^{\text{val}}_\beta = e\cdot
\frac{\frac{D m_{\text{DOS}}^\text{LH}\tau^\text{LH}_\beta}{(m^*_\text{LH})_\beta} + \frac{D
  m_{\text{DOS}}^\text{HH}\tau^\text{HH}_\beta}{(m^*_\text{HH})_\beta}}{D(m_{\text{DOS}}^\text{LH} +
m_{\text{DOS}}^\text{HH})}=  
e\cdot
\frac{\frac{m_{\text{DOS}}^\text{LH}\tau^\text{LH}_\beta}{(m^*_\text{LH})_\beta} + \frac{
  m_{\text{DOS}}^\text{HH}\tau^\text{HH}_\beta}{(m^*_\text{HH})_\beta}}{m_{\text{DOS}}^\text{v}}
\end{equation}
where $m_{\text{DOS}}^\text{v}$ is the total DOS mass in the valence. Finally, we
can prove that in the valence the mobility does not depend on the DOS
mass of each of the two valleys, but only on the total DOS mass and on the
effective masses in the transport direction. In fact, by simply
replacing Supplementary Equation~\eqref{eq:oneovertau}, one finally obtains:
\begin{equation}
\label{eq:mobility-val-nosoc}\footnotesize
\boxed{
\mu^{\text{val}}_{\beta}[\text{without SOC}] = \frac{e\hbar^3C_\beta}{k_\text{B} T}\cdot
\frac{\frac{1}{(E^\text{LH}_\beta)^2(m^*_\text{LH})_{\beta}} + 
\frac{1}{(E^\text{HH}_\beta)^2(m^*_\text{HH})_\beta}}
{m_{\text{DOS}}^\text{v}}}.
\end{equation}

When we include spin-orbit effects, we need to consider only the
topmost valence band (as already discussed), and since the band is
isotropic, we can also set its mass $m^*_\text{v} = m_{\text{DOS}}^\text{v}$. The formula then
simplifies to:
\begin{equation}
\label{eq:mobility-val-soc}
\boxed{
\mu^{\text{val}}_{\beta}[\text{with SOC}] = \frac{e\hbar^3C_\beta}{k_\text{B} T}\cdot
\frac{1}{(E^\text{LH}_\beta)^2(m^*_\text{v})^2}}.
\end{equation}

\subsection{Calculation of the deformation potential}

\begin{figure}[tb]
  \centering
  \includegraphics[width=0.45\linewidth]{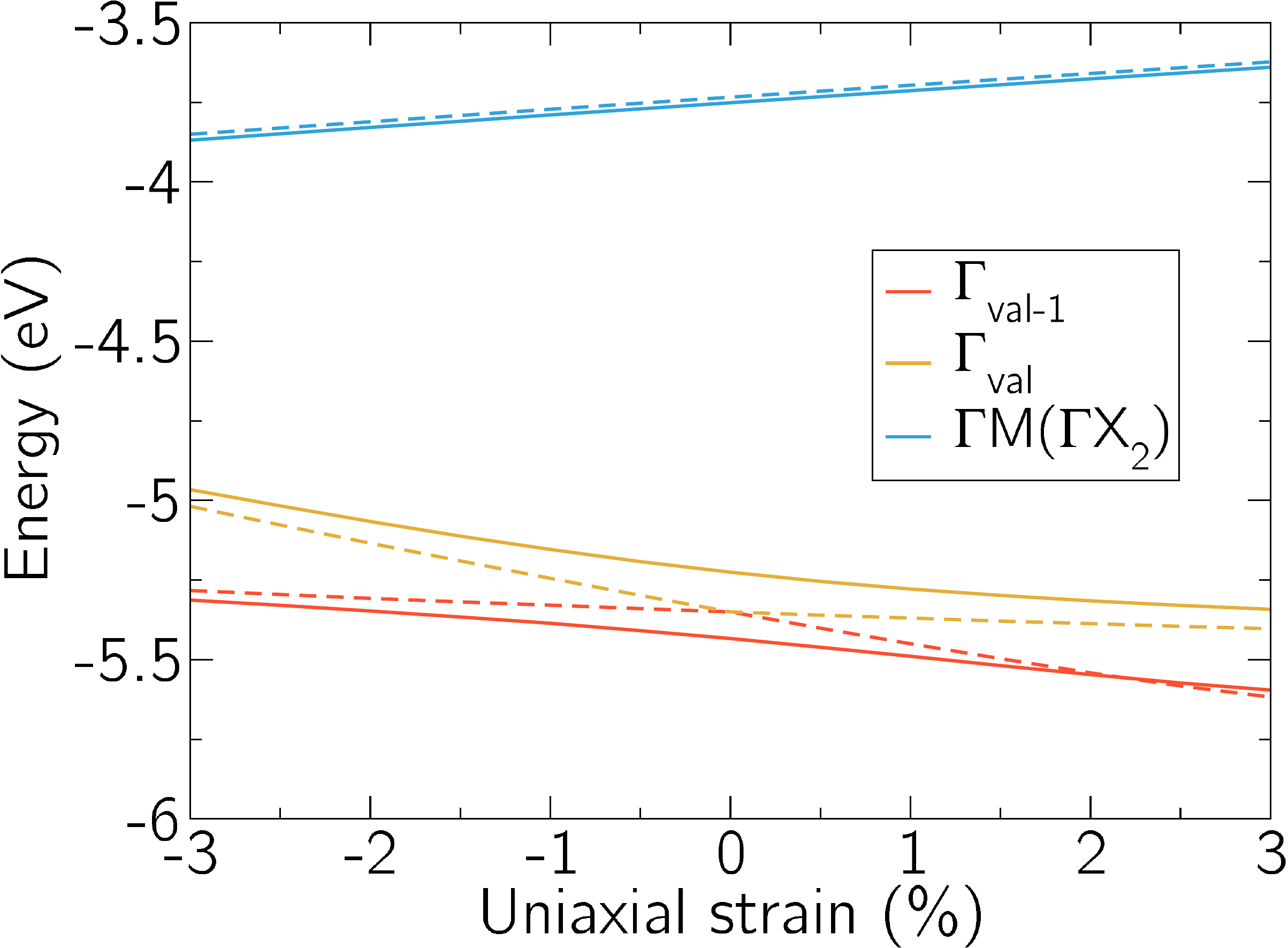}\hfill%
  \includegraphics[width=0.45\linewidth]{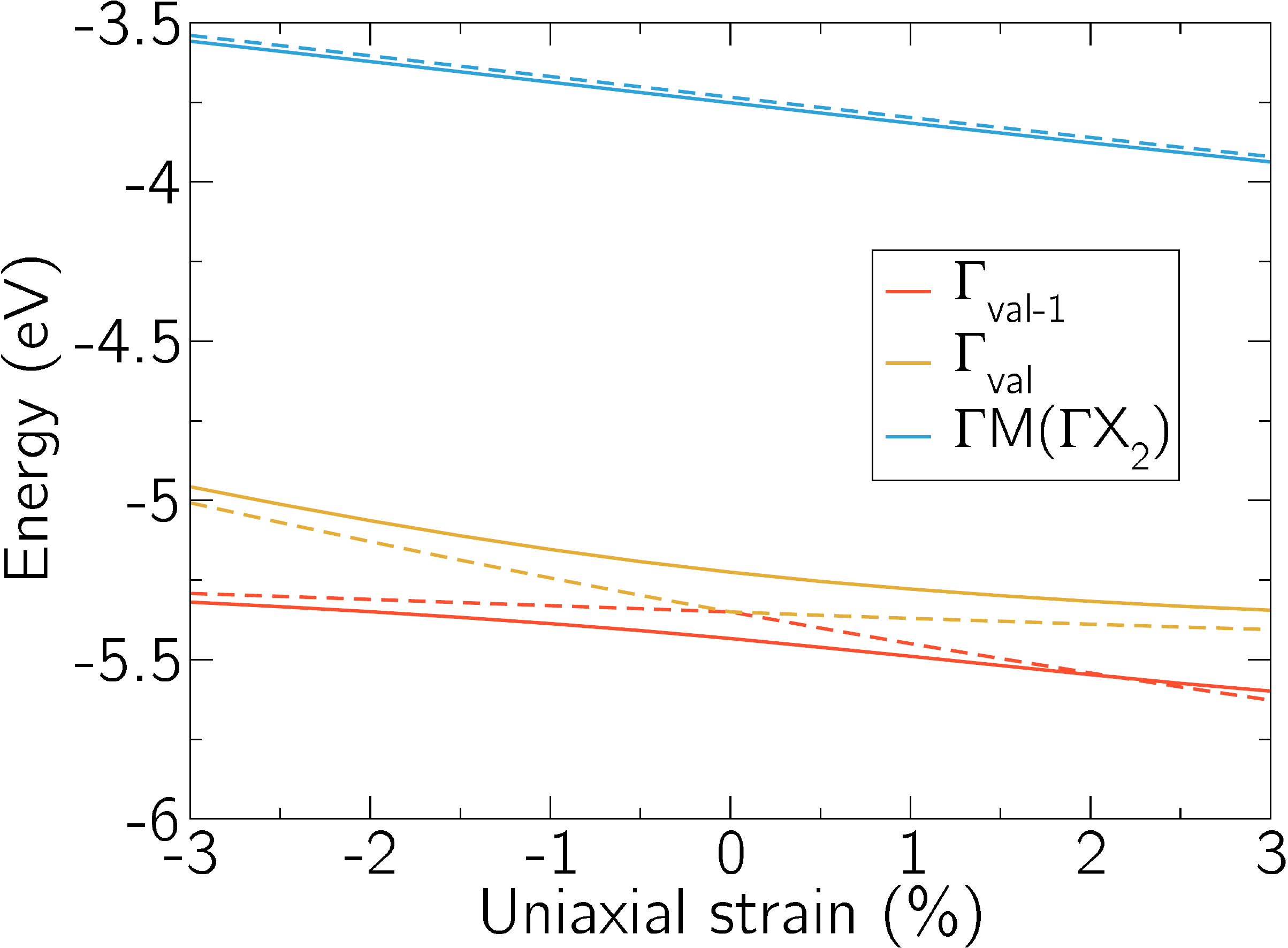}
  \caption{Vacuum-level-corrected band edges as a function of the strain for the
As monolayers. Left panel: strain along the $\vec w_1$ direction (armchair).
Right panel: strain along the $\vec w_2$ direction (zigzag). 
$\Gamma_\text{val}$ indicates the topmost valence band at $\Gamma$, 
$\Gamma_\text{val-1}$ indicates the second valence band. Dashed curves refer to the case without SOC, while solid curves include SOC effects.
  \label{fig:as-bandedges}}
\end{figure}

\begin{figure}[tb]
  \centering
  \includegraphics[width=0.45\linewidth]{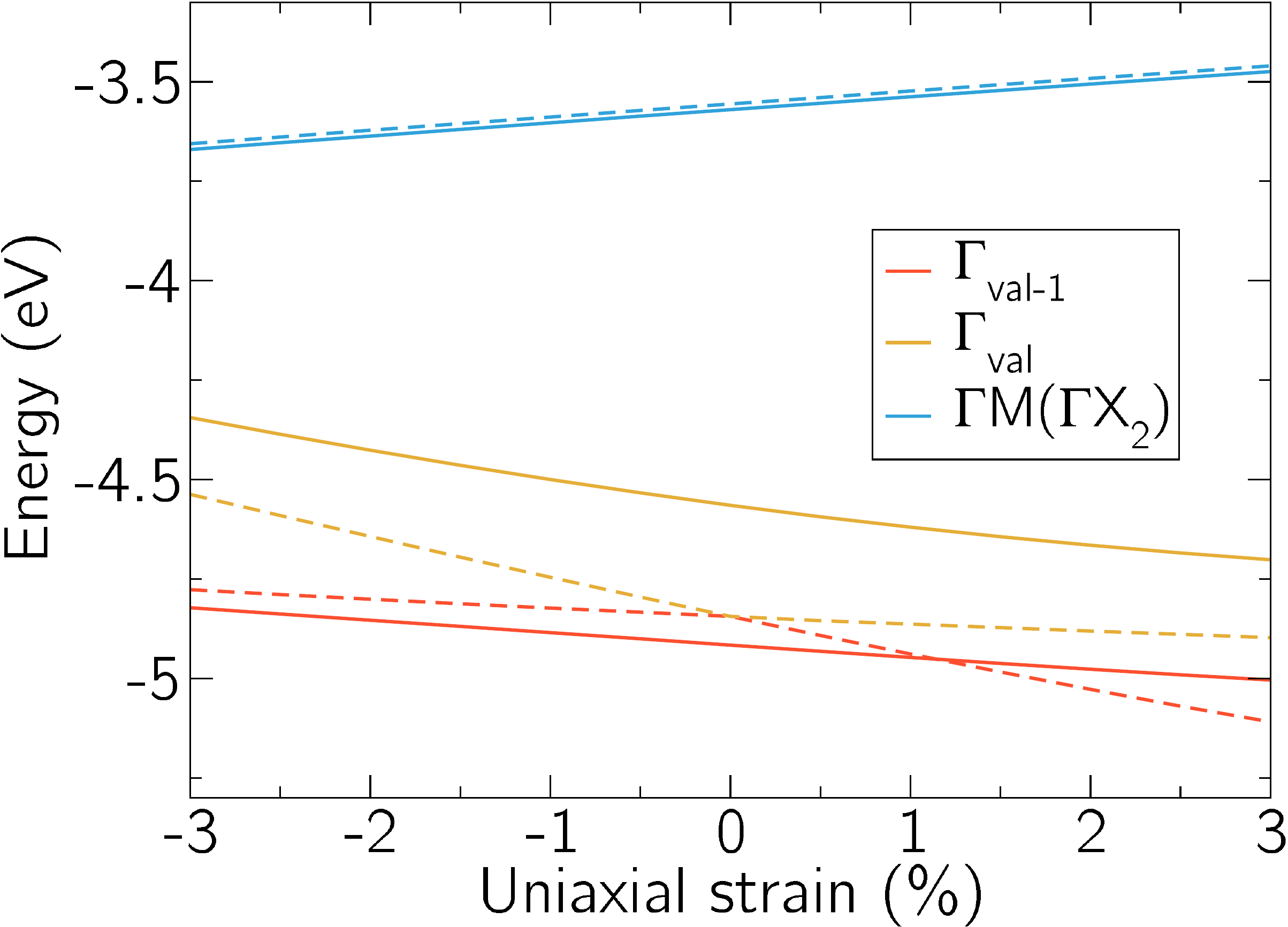}\hfill%
  \includegraphics[width=0.45\linewidth]{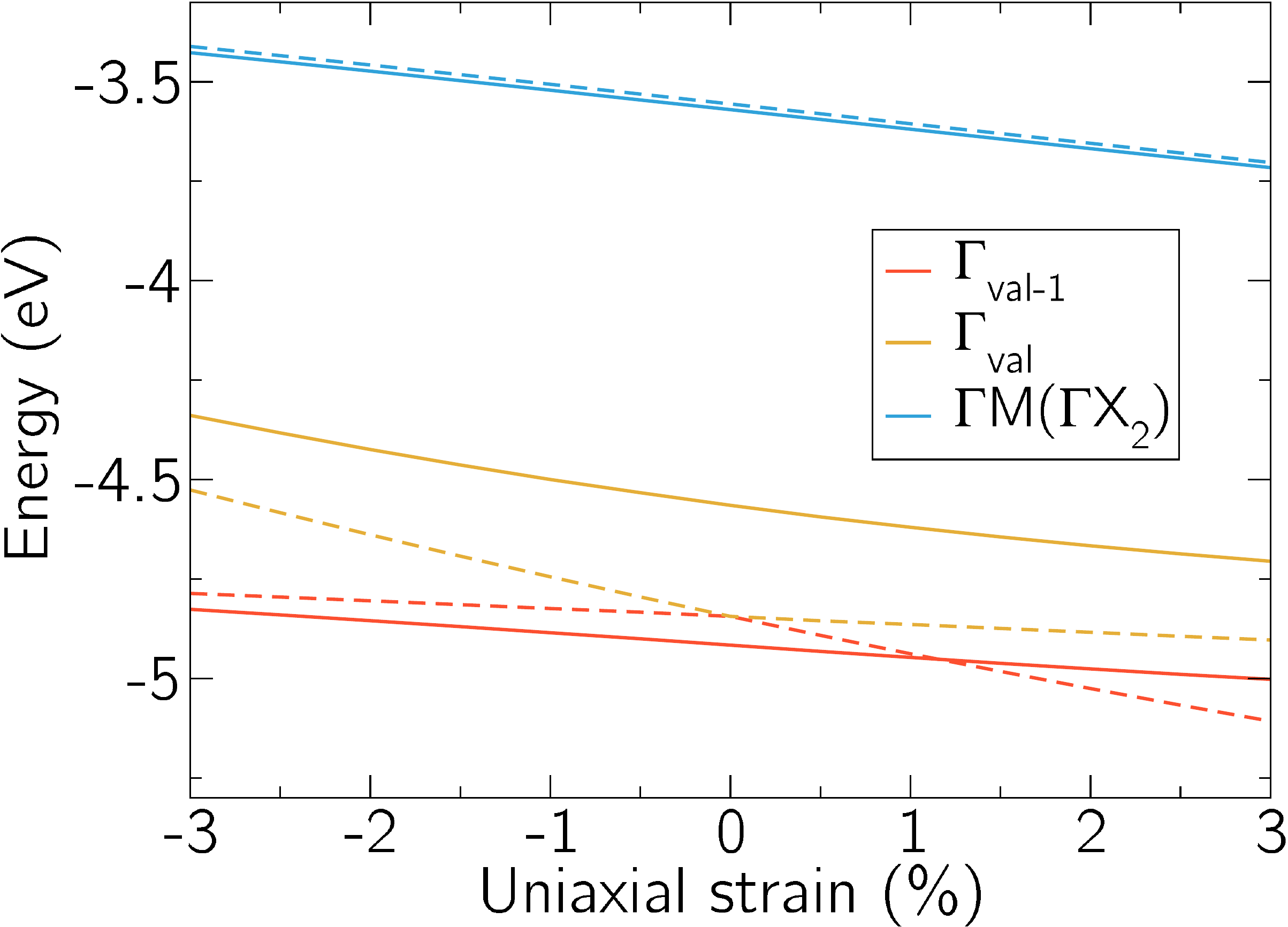}
  \caption{Vacuum-level-corrected band edges as a function of the strain for the
Sb monolayers. Left panel: strain along the $\vec w_1$ direction (armchair).
Right panel: strain along the $\vec w_2$ direction (zigzag). 
$\Gamma_\text{val}$ indicates the topmost valence band at $\Gamma$, 
$\Gamma_\text{val-1}$ indicates the second valence band. Dashed curves refer to the case without SOC, while solid curves include SOC effects.
  \label{fig:sb-bandedges}}
\end{figure}

The deformation potential is calculated using finite differences
starting from band energies calculated with DFT
of systems strained along the two relevant
transport directions (zigzag and armchair). 
Considering a given band edge $i$ (for instance the
topmost valence band at $\Gamma$, or the conduction band along the
$\Gamma-$M line in 
As and Sb), we can define $\Delta V_\beta^i= \eps^i(\Delta l_\beta) -
\eps^i(\Delta l_\beta = 0)$, where $\eps^i(\Delta l_\beta)$ is the 
energy of the $i-$th band edge for a system strained in the
$\beta$ direction by a quantity $\Delta l_\beta$. The deformation potential is then
simply $E^i_\beta=\Delta V_\beta^i / (\Delta l_\beta/l_{\beta,0})$,
where $l_{\beta,0}$ is the relaxed value of the lattice constant in
the $\beta$ direction, in the limit of small strains $\Delta l_\beta$. 
It is important to stress that the band-edge energies directly
extracted from DFT calculations of different systems are ill-defined, 
because the position of the vacuum level can change in each
calculation. We therefore use always vacuum-level-corrected band
energies $\eps^i$, obtained by defining, at each strain $\beta$, the vacuum level 
as the zero of energy. The vacuum level is obtained by
calculating the averaged electrostatic potential in the region of space 
far away from the 2D layers. We verified that in all cases such potential is flat 
(within a 0.01 meV precision) $4-5$~\AA{} away from the monolayers.
We also note that the primitive 2D cell (with 2 atoms per cell) 
has non-orthogonal lattice vectors. In order to define strained
systems in the two transport directions, we define a rectangular cell
with 4 atoms per cell, where the two lattice vectors $\vec w_1$ and $\vec w_2$
are in the armchair and zigzag direction respectively, as shown in 
Supplementary Figure~\ref{fig:realspaceandBZcell-2xsupercell}.
Uniaxial strains are then applied to this rectangular supercell.

We consider 13 calculations for different, uniformly spaced strains
between $-3\%$ and
$3\%$ for each of the two $\beta$ directions,
and we extract the deformation potential from the linear coefficient of
a quadratic fit of the vacuum-level-corrected band 
edges, as shown in Supplementary 
Figs.~\ref{fig:as-bandedges} and~\ref{fig:sb-bandedges}, where we report
results both with and without SOC.

\subsection{Effective masses}
\label{sec:masses}
\begin{table}[tb]
\caption{\label{tab:defpot}
Deformation potentials (in eV) for the valence and conduction bands of As and Sb 
monolayers, both including and disregarding spin-orbit coupling effects. For valence bands with spin-orbit coupling, we only report the deformation potential of the topmost valence band; in the case without spin-orbit coupling, $i=1,2$ are defined so that
$i=1$ is the topmost valence band for positive
strain, and $i=2$ the second highest valence band for positive strain,
equivalent to saying that we chose $i=1,2$ so that
$E^{\text{val}}_\beta(i=2) < E^{\text{val}}_\beta(i=1)$. Results without spin-orbit coupling are consistent with previous calculations at the same level of theory~\cite{Wang2015}.}
\begin{center}
\footnotesize
\begin{tabular}{l..}
  With SOC& \multicolumn{1}{c}{$E^{\text{val}}_\beta$} &
    \multicolumn{1}{c}{$E^{\text{cond}}_\beta(\Gamma-\text{X})$} 
  \\\hline\hline
As ($\beta = $ armchair) & $-6.243$ & $3.815$
\\ 
As ($\beta = $ zigzag)   & $-6.386$ & $-6.351$ 
\\\hline 
Sb ($\beta = $ armchair) & $-5.966$ & $3.265$
\\ 
Sb ($\beta = $ zigzag)   & $-6.078$ & $-4.843$
\\ 
\end{tabular}
\vspace{2ex}
\begin{tabular}{l...}
  Without SOC& \multicolumn{1}{c}{$E^{\text{val}}_\beta(i=1)$} &
    \multicolumn{1}{c}{$E^{\text{val}}_\beta(i=2)$}  &
    \multicolumn{1}{c}{$E^{\text{cond}}_\beta(\Gamma-\text{X})$} 
  \\\hline\hline
As ($\beta = $ armchair)  & $-1.997$ & $-10.094$ & $3.785$ 
\\ 
As ($\beta = $ zigzag)  & $-1.922$ & $-10.321$ &$-6.380$ 
\\\hline 
Sb ($\beta = $ armchair)  &$-2.006$ & $-9.571$ & $3.260$
\\ 
Sb ($\beta = $ zigzag)  & $-1.969$ & $-9.674$ & $-4.895$
\\ 
\end{tabular}
\end{center}
\end{table}

We have computed the effective masses at zero strain in the primitive unit cell; we have
checked that the effective masses do not change significantly with strain. The values (calculated as described below) are reported in Table~1 of the main paper when including SOC effects. We also report here, in Supplementary Table~\ref{tab:effmasses-nosoc}, the values calculated when SOC is not included.

Also in the case of effective masses, we have to distinguish two cases.
In the conduction band, 
the masses can be calculated by a parabolic fit of the band energies along the two directions. We have used for the fit
 21 $\vec k-$points around the band minimum (at zero strain) in the longitudinal and transverse directions.
  The DOS mass $m_{\text{DOS}}^\text{c}$ is instead obtained by a parabolic fit of the integrated DOS (to take into 
account non-parabolicity effects). The DOS has been calculated on a dense $\vec k-$mesh using Wannier interpolation.
  The conduction valleys are in a very good approximation parabolic and indeed
  $m_{\text{DOS}}^\text{c}\approx \sqrt{m^\text{c}_\text{L}m^\text{c}_\text{T}}$ (see Table~1 in the main paper).

In the valence band, without SOC and in the absence of strain
we have two isotropic bands with two different masses, named heavy hole (HH) and
light hole (LH) for the larger and smaller mass (in absolute value), 
respectively.
A difficulty arises when we apply some strain: in this case, the two bands
split and we need to know how to associate the lower-energy and higher-energy
valence bands to corresponding effective mass (LH or HH).
Actually, with strain the two masses become anisotropic tensors.
In particular, given a direction $\beta$ for the strain, 
we have checked that (both for As and Sb) the effective mass obtained
fitting the bands for $\vec k-$points in the same direction as $\beta$ is HH for the
bands with larger deformation potential (smaller in absolute value, since $E<0$), while LH for the band with smaller deformation potential.
Instead, if the $\vec k-$points are taken in a direction orthogonal to $\beta$, the two masses HH and LH are reversed.
Since in Supplementary Equation~\eqref{eq:mobility-val-nosoc} we need only the deformation
potential for strains along the same transport direction as the 
effective mass components, Supplementary Equation~\eqref{eq:mobility-val-nosoc} becomes:
\begin{equation}
\mu^{\text{val}}_{\beta} = \frac{e\hbar^3C_\beta}{k_\text{B} T}\cdot
\frac{\frac{1}{(E_\beta(v,i=1))^2(m^*_\text{HH})_{\beta}} + 
\frac{1}{(E_\beta(\text{v},i=2))^2(m^*_\text{LH})_{\beta}}}
{m_{\text{DOS}}^\text{v}},
\end{equation}
where we have defined $i=1,2$ consistently to 
Supplementary Table~\ref{tab:defpot}, i.e., so
that $E_\beta(\text{v},i=2) < E_\beta(\text{v},i=1)$.

In the case with SOC, this difficulty does not arise because the two 
bands do not cross.

\subsection{Elastic moduli}
\begin{figure}[tb]
  \centering
  \includegraphics[width=0.45\linewidth]{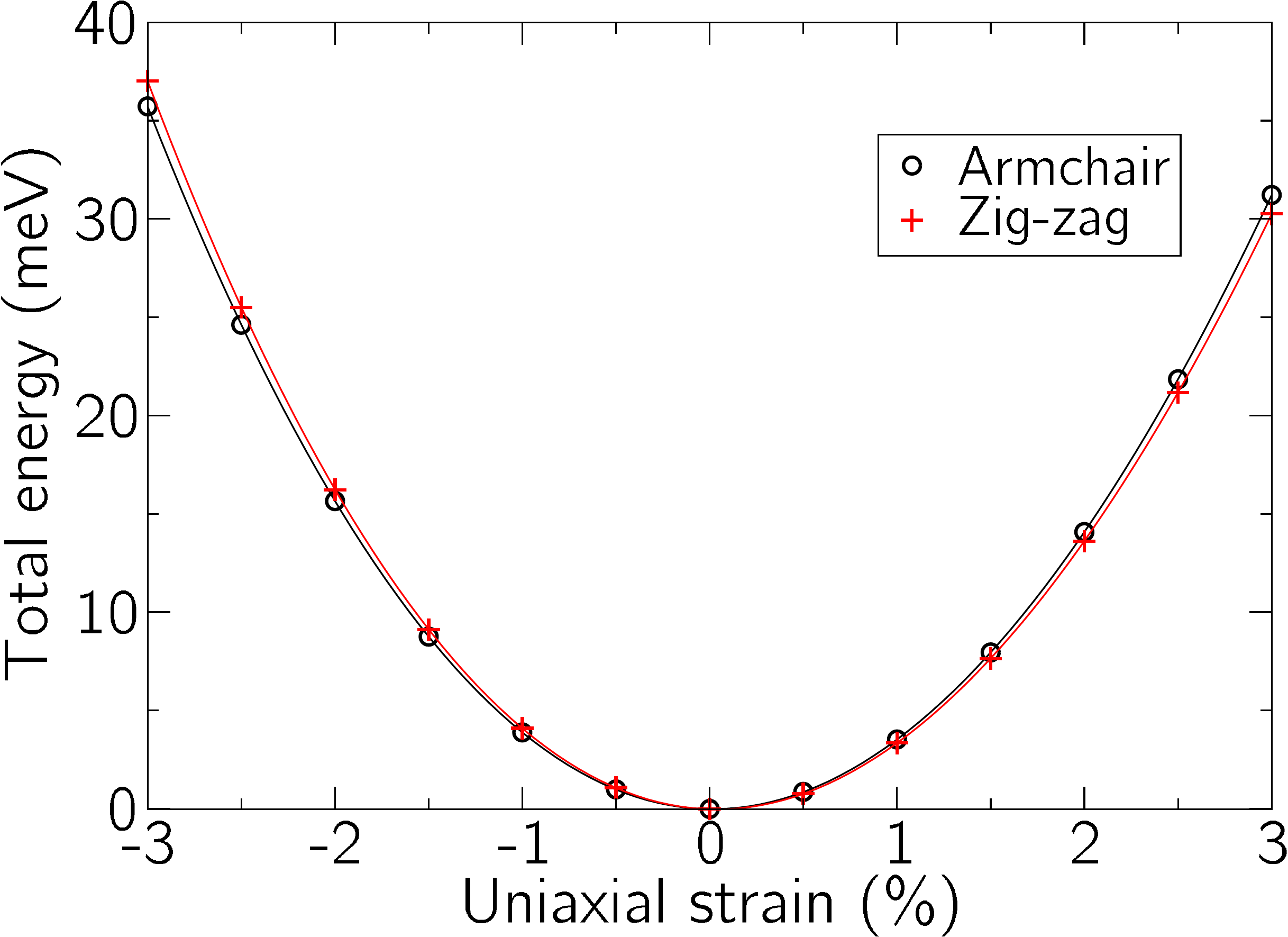}\hfill%
  \includegraphics[width=0.45\linewidth]{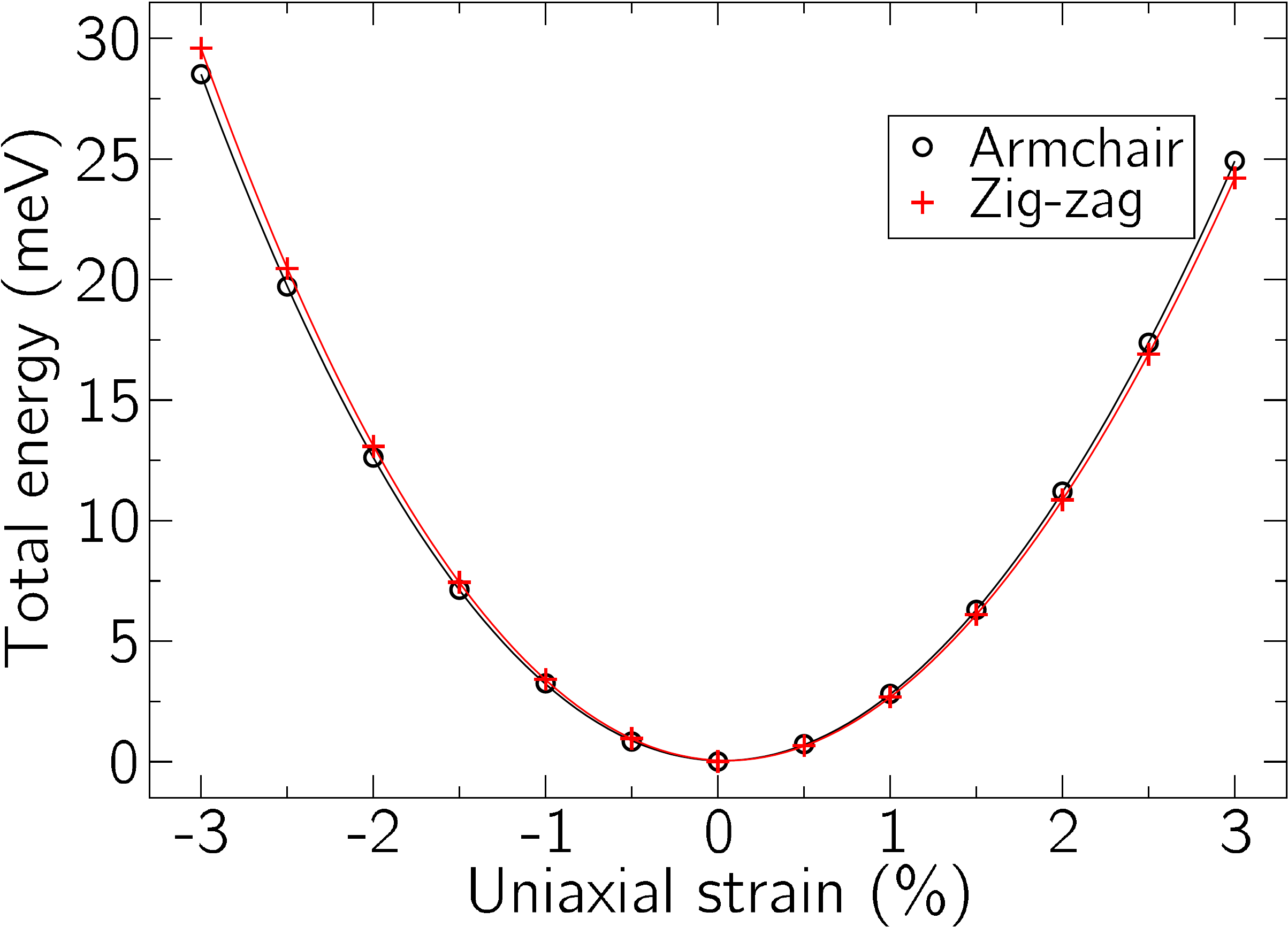}
  \caption{Total energy as a function of the uniaxial strain (when SOC effects are not included) in the armchair  (open points) and zigzag (plus symbols) direction.   Left panel: As monolayer; right panel: Sb monolayer.
    The curves are the cubic fits of the data points used to obtain the
    elastic moduli.
  \label{fig:uniaxialeos}}
\end{figure}
\begin{table}[tb]
\footnotesize
\caption{\label{tab:elasticmoduli}
2D elastic moduli of the As and Sb monolayer systems, both including and disregarding spin-orbit coupling effects.}
\begin{tabular}{lccc}
 & \multicolumn{1}{c}{$S_0$ (\AA$^2$)} &
 \multicolumn{1}{c}{$C_\beta$ (eV\,\AA$^{-2}$)} &
 \multicolumn{1}{c}{$C_\beta$ (eV\,\AA$^{-2}$)} \\
 & &  \tiny with SOC & \tiny without SOC \\\hline\hline
As ($\beta = $ armchair) & \multirow{ 2}{*}{22.5} & 3.271 & 3.299 \\ 
As ($\beta = $ zigzag) & & 3.279 & 3.308 \\\hline 
Sb ($\beta = $ armchair) &  \multirow{ 2}{*}{29.4} & 1.948 & 2.012 \\ 
Sb ($\beta = $ zigzag) & & 1.945 & 2.019 \\ 
\end{tabular}
\end{table}

The 2D elastic modulus in a given direction $\beta$ is
defined as:
\begin{equation}
C_\beta=\frac 1 {S_0}\left.\frac{\partial^2 \mathcal{E}_{\Delta
    l_\beta}}{\partial(\Delta l_\beta/l_{\beta,0})^2}\right|_{\Delta l_\beta
= 0},
\end{equation}
where $\mathcal{E}_{\Delta
    l_\beta}$ is the total energy of a system strained uniaxially 
  by $\Delta l_\beta/l_{\beta,0}$ in
  the $\beta$ direction, and $S_0$ is the value of
  the unstrained unit-cell surface.

The values reported in the paper are obtained from the parabolic coefficient of
a cubic fit of the total energy as a function of strain in the $[-3,+3]\%$ 
range for the four-atom supercells, shown in Supplementary Figure~\ref{fig:uniaxialeos} for both
materials (when SOC is included). In a hexagonal system the elastic tensor should be isotropic in the
plane. Indeed, apart from numerical inaccuracies, the $C_\beta$ values are the same for both directions (see Supplementary Table~\ref{tab:elasticmoduli}), and do not significantly change including or disregarding SOC.

\subsection{Values of the mobility}
\begin{table}[tb]
\caption{\label{tab:effmasses-nosoc} Effective masses of the relevant bands
  of arsenene and antimonene, in units of the electron mass $m_0$, 
  when SOC is not included.
  Symbols are explained in the text.  Note that
  $m^{\text{c}}_{\text{DOS}}$ is the effective DOS mass for each of the
  6 identical conduction band valleys, while $m^{\text{v}}_{\text{DOS}}$
  represents the total effective DOS mass for the two
  degenerate valence bands.}
\footnotesize
\begin{tabular}{lcc}
 & As & Sb \\\hline\hline $m^{\text{c}}_{\text{DOS}}$ & 0.273 & 0.260
  \\ $m^{\text{c}}_\text{L}$ & 0.508 & 0.461 \\ $m^{\text{c}}_\text{T}$ & 0.150 & 0.149 \\
$m^{\text{v}}_{\text{DOS}}$ & 0.554 & 0.523 \\ $m^{\text{v}}_{\text{HH}}$ & 0.482 &
  0.443 \\ $m^{\text{v}}_{\text{LH}}$ & 0.077 & 0.073 \\
\end{tabular}
\end{table}
\begin{table}[tb]
\caption{\label{tab:mobility-final} 
Mobilities for arsenene and antimonene, estimated at $T=300\,\text{K}$, in units of $\text{cm}^2\text{V}^{-1}\text{s}^{-1}$. Both the values obtained with and without spin-orbit coupling (SOC) are reported.
}
\hfil\begin{tabular}{cp{0.5ex}ccp{0.5ex}cc}
&& \multicolumn{2}{c}{With SOC} && \multicolumn{2}{c}{Without SOC}
\\
 && \multicolumn{1}{c}{$\mu^{\text{val}}$} 
 & \multicolumn{1}{c}{$\mu^{\text{cond}}$} &
& \multicolumn{1}{c}{$\mu^{\text{val}}$} 
 & \multicolumn{1}{c}{$\mu^{\text{cond}}$}\\\hline\hline
As && 1700 & 635 && 1355 & 622 \\
Sb && 1737 & 630 && 946 & 641 \\
\end{tabular}\hfil
\end{table}

Using the values of Supplementary Table~\ref{tab:defpot}, Table~1 (in the main paper) and Supplementary Table~\ref{tab:elasticmoduli}, calculated as described in the previous sections, we estimated the values for the mobilities at $T=300\,\text{K}$, in units of $\text{cm}^2\text{V}^{-1}\text{s}^{-1}$, that we report in Supplementary Table~\ref{tab:mobility-final}.
Note that since we expect that the total mobility is isotropic, we indicate only the average value obtained for the armchair and the zigzag directions. In any case, the values in the two directions differ (due to numerical errors) by less than 5\%.

For completeness, we also report the value of the single-valley conduction mobility tensor, including spin-orbit coupling effects, in a basis set
where the first vector is along the armchair direction and 
the second along the zigzag one, and expressed in units of $\text{cm}^2\text{V}^{-1}\text{s}^{-1}$:
\begin{equation}
  \label{eq:mobility-cond-singlevalley} 
  \mu^{\text{cond},i}_{\text{As}} = 
  \begin{pmatrix}
    567 & 0 \\
    0 & 704
  \end{pmatrix}, \qquad
  \mu^{\text{cond},i}_{\text{Sb}} = 
  \begin{pmatrix}
    507 & 0 \\
    0 & 753
  \end{pmatrix}.
\end{equation}

Using the formulas above, one can also calculate the scattering times of electrons with LA phonons, that turn out to be (in conduction): $\tau_{\text{armchair}}^{\text{As}} = {161}\,\text{fs}$, $\tau_{\text{zigzag}}^{\text{As}} = {58}\,\text{fs}$, $\tau_{\text{armchair}}^{\text{Sb}} = {136}\,\text{fs}$, $\tau_{\text{zigzag}}^{\text{Sb}} = {62}\,\text{fs}$. Finally, in order to estimate the mean free path, it is useful to evaluate the Fermi velocity for electrons in the $\Gamma-$M valleys, that for electrons along the $i$ principal direction of the valley  is given by:
\begin{equation}
v^\text{F}_i = \sqrt{\frac{2\pi\hbar^2n_{\text{dop}}}{m_im_{\text{DOS}}}},
\end{equation}
where we have used the 2D DOS to relate the the effective doping of the system $n_{\text{dop}}$ to the Fermi energy, and where $m_i$ indicates the effective mass
in the $i$ direction. As a reference value, for a doping of $n_{\text{DOP}}=5\cdot 10^{13}\,\text{cm}^{-2}$, we obtain for As:  $v^\text{F}_{\text{L}} = {0.55}\,\text{nm\,fs}^{-1}$, $v^\text{F}_{\text{T}} = {1.01}\,\text{nm\,fs}^{-1}$, while for Sb: $v^\text{F}_{\text{L}} = {0.60}\,\text{nm\,fs}^{-1}$, $v^\text{F}_{\text{T}} = {1.04}\,\text{nm\,fs}^{-1}$.

\subsection{n-MOS and p-MOS Field Effect Transistors}
\begin{figure}[tb]
  \centering
  \includegraphics[width=0.85\linewidth]{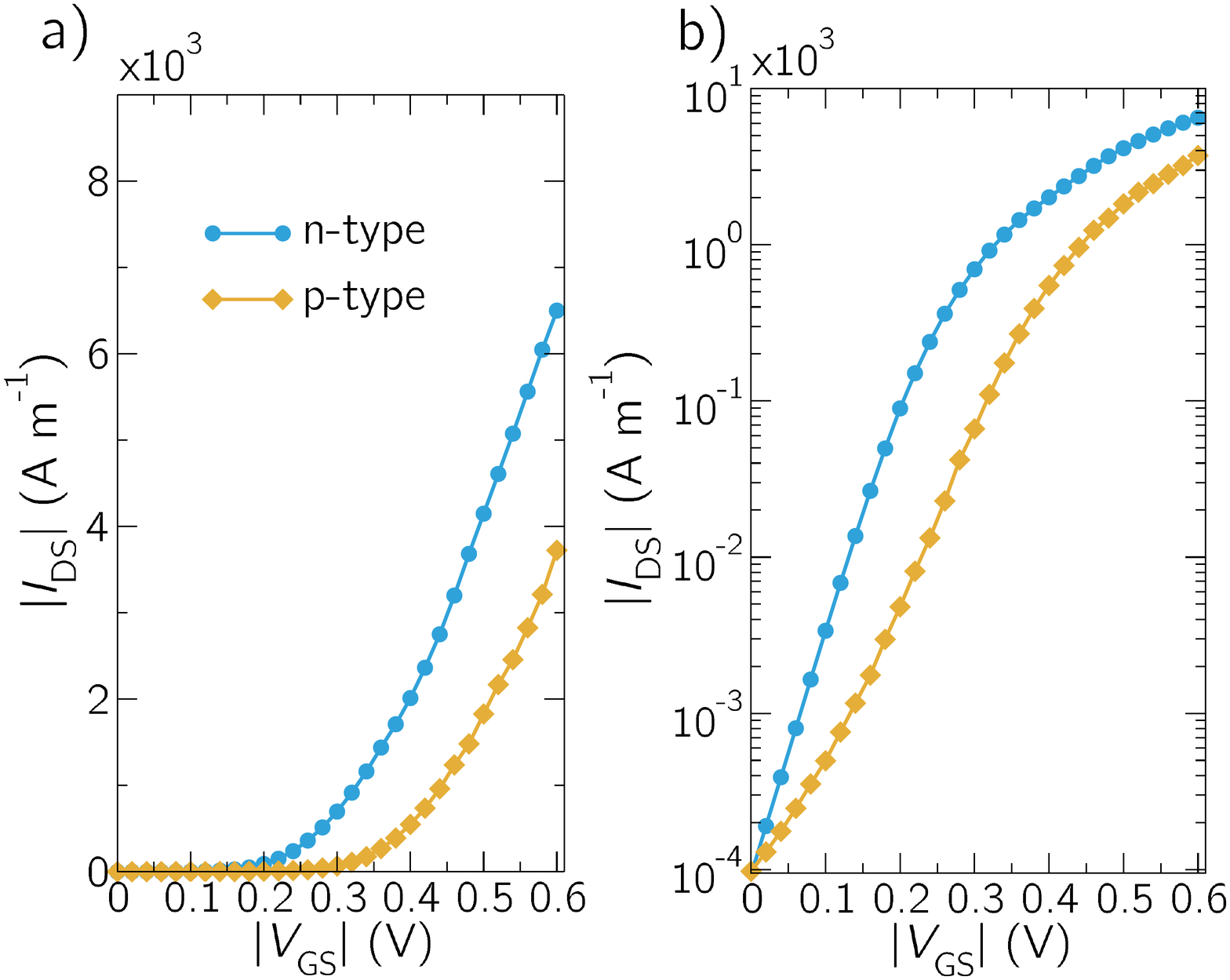}
  \caption{$|I_\text{DS}|-|V_\text{GS}|$ curve in (a) linear and (b) semi-logarithmic scale
    for As n-MOS and p-MOS transistors and a gate length of $L_\text{G}$=7~nm.
    Sb FETs show a similar behaviour and are not shown here. \label{fig:nMOSpMOS}}
\end{figure}

In order to span the whole device parameter space and optimise the device performances, we compare the $I-V$ curves of 
n-MOS and p-MOS devices in Supplementary Figure~\ref{fig:nMOSpMOS}, in the
case of a As device with $L_\text{G}=7$~nm (similar results are obtained for Sb FETs).

As it can be seen, the n-MOS device shows larger currents and better SS
as compared to the p-type device. For this reason, in the main text
we decided to focus on n-MOSFETs only,
in order to obtain the best performance against Industry requirements.


\end{document}